\documentclass[fleqn,usenatbib]{mnras}

\usepackage{newtxtext,newtxmath}
\usepackage[T1]{fontenc}
\usepackage{ae,aecompl}
\usepackage{times}
\usepackage{enumitem}
\usepackage{times}
\usepackage{textcomp}
\usepackage{verbatim}
\usepackage{esdiff}

\newcommand{\msun}{{\,\rm M_\odot}}

\newcommand{\kms}{\,{\rm km}\,{\rm s}^{-1}}

\newcommand{\Gyr}{\,{\rm Gyr}}

\newcommand{\pc}{\,{\rm pc}}
\newcommand{\kpc}{\,{\rm kpc}}

\newcommand{\cpm}{\,{\rm cm}^2\,{\rm g}^{-1}}

\newcommand{\mmag}{\,{\rm mag}}

\def\jcap{J. Cosmol.  Astropart. Phys.}
\def\aap{A\&A}
\def\apj{ApJ}

\def\apjl{ApJ}
\def\mnras{MNRAS}
\def\araa{ARA\&A}
\def\aj{AJ}

\def\physrep{Phys. Rep.}
\def\nat{Nature}

\def\apjs{ApJS}
\def\prd{Phys. Rev. D}

\usepackage{graphicx}	% Including figure files
\usepackage{amsmath}	% Advanced maths commands

\usepackage{tikz}
\usepackage{adjustbox}
\makeatletter
\newcommand{\rmnum}[1]{\romannumeral #1}
\newcommand{\Rmnum}[1]{\expandafter\@slowromancap\romannumeral #1@}

\renewcommand\paragraph{\@startsection{paragraph}{4}{\z@}{3.25ex\@plus1ex\@minus.2ex}{-1em}{\normalfont\it\normalsize}}

\defcitealias{Shen2021}{Paper~\Rmnum{1}}

\title[Dissipative Dark Matter on FIRE]{Dissipative Dark Matter on FIRE: \Rmnum{2}. Observational signatures and constraints from local dwarf galaxies}

\author[Shen et al.]{\parbox{17.0cm}{
Xuejian Shen$^{1}$\thanks{Contact e-mail: \href{mailto:xshen@caltech.edu}{xshen@caltech.edu}},
Philip F. Hopkins$^{1}$,
Lina Necib$^{2,3}$,
Fangzhou Jiang$^{1,4}$,
Michael Boylan-Kolchin$^{5}$,
Andrew Wetzel$^{6}$
\\
}\vspace{0.3cm}\\
$^{1}$TAPIR, California Institute of Technology, Pasadena, CA 91125, USA\\
$^{2}$ Kavli Institute for Astrophysics and Space Research, Massachusetts Institute of Technology, 77 Massachusetts Ave, Cambridge MA 02139, USA\\
$^{3}$ The NSF AI Institute for Artificial Intelligence and Fundamental Interactions\\
$^{4}$Carnegie Observatories, 813 Santa Barbara Street, Pasadena, CA 91101, USA\\
$^{5}$Department of Astronomy, The University of Texas at Austin, 2515 Speedway Stop C1400, Austin, TX 78712, USA\\
$^{6}$Department of Physics \& Astronomy, University of California, Davis, CA 95616, USA
}

\date{Accepted XXX. Received YYY; in original form ZZZ}

\pubyear{2022}

\begin{document}
\label{firstpage}
\pagerange{\pageref{firstpage}--\pageref{lastpage}}
\maketitle

\begin{abstract}
We analyze the first set of cosmological baryonic zoom-in simulations of galaxies in dissipative self-interacting dark matter (dSIDM). The simulations utilize the FIRE-2 galaxy formation physics with the inclusion of dissipative dark matter self-interactions modelled as a constant fractional energy dissipation ($f_{\rm diss}=0.5$). In this paper, we examine the properties of dwarf galaxies with $M_{\ast} \sim 10^{5}\operatorname{-}10^{9}\msun$ in both isolation and within Milky Way-mass hosts. For isolated dwarfs, we find more compact galaxy sizes and promotion of stellar/neutral gas disk formation in dSIDM with $(\sigma/m)\leq 1\cpm$ but they are still consistent with observed galaxy sizes and masses. On contrary, models with $(\sigma/m) = 10\cpm$ produces puffier stellar distributions that lie in the diffuse end of the observed size-mass relation. In addition, as a result of the steeper central density profiles developed in dSIDM, the sub-kpc circular velocities of isolated dwarfs in models with $(\sigma/m)\geq 0.1\cpm$ are enhanced by about a factor of two, which are still consistent with the measured stellar velocity dispersions of Local Group dwarfs but in tension with the H\Rmnum{1} rotation curves of more massive field dwarfs. Meanwhile, for satellites of the simulated Milky Way-mass hosts, the median circular velocity profiles are marginally affected by dSIDM physics, but dSIDM may help address the missing compact dwarf satellites in CDM. The number of satellites is slightly enhanced in dSIDM, but the differences are small compared with the large host-to-host variations revealed in observations. In conclusion, the dSIDM models with constant cross-section $(\sigma/m) \gtrsim 0.1\,{\rm cm^{2}\,g^{-1}}$ (assuming $f_{\rm diss}=0.5$) are effectively ruled out in bright dwarfs ($M_{\rm halo}\sim 10^{11}\msun$) by circular velocity constraints. However, models with lower effective cross-sections (at this halo mass/velocity scale) are still viable and can give rise to non-trivial observable signatures. 
\end{abstract}


%%%%%%%%%%%%%%%%%%%%%%%%%%%%%%%%%%%%%%%%%%%%%%%%%%

%%%%%%%%%%%%%%%%% BODY OF PAPER %%%%%%%%%%%%%%%%%%
\section{Introduction}

Although dark matter is widely recognized as an essential component of the Universe and the main driver for cosmological structure formation, its nature remains a mystery. In the concordance model of cosmology -- the cosmological constant plus cold dark matter ($\Lambda {\rm CDM}$) model, dark matter is assumed to be non-relativistic (``cold'') at decoupling from the initial plasma and effectively collisionless. This picture has been successful in describing the large-scale structures in the Universe~\citep[e.g.,][]{Blumenthal1984,Davis1985} and form the foundation for theories of galaxy formation and evolution~\citep[e.g.,][]{White1991,Kauffmann1993,Cole2000}.

However, some aspects of the classical CDM scenario have been challenged by the lack of signal from direct detection experiments and apparent anomalies found in astrophysical observations, which have motivated conjectures of alternative dark matter models. One of the most popular candidates for CDM (the class of Weakly Interacting Massive Particles, WIMPs) has not been detected in collider searches and direct or indirect detection experiments. With a significant proportion of the supersymmetric WIMP parameter space being ruled out~\citep[e.g.,][]{Bertone2005,Bertone2010,Aprile2012,Akerib2014,Aprile2018}, the null results in these experimental searches stimulate thoughts about alternative models for dark matter~\citep[e.g.,][]{Hogan2000,Spergel2000,Dalcanton2001,Buckley2018}. Meanwhile, the $\Lambda {\rm CDM}$ model still faces significant challenges in matching astrophysical observations at small scales~\citep[see the review][]{Bullock2017} and particularly in dwarf galaxies. In the past, the classic example of this was the {\it missing satellites} (MS) problem stems from the overproduction of dark matter subhaloes around Milky Way-mass hosts in dark-matter-only (DMO) simulations compared to the observed luminous satellites of the real Milky Way \citep[e.g.,][]{Klypin1999, Moore1999}. Another small-scale question, the {\it core-cusp} problem states that the central density profiles of dark matter dominated systems, e.g. low-surface-brightness galaxies (LSBs) and dwarf spheroidal galaxies (dSphs), are cored~\citep[e.g.,][]{Flores1994,Moore1994,deBlok2001,KDN2006,Gentile2004,Simon2005,Spano2008,KDN2011a,KDN2011b,Oh2011,Walker2011,Oh2015,Chan2015,Zhu2016}, in contrast to the universal cuspy central profile found in DMO simulations~\citep{Navarro1996,Navarro1997,Moore1999,Klypin2001,Navarro2004,Diemand2005}. Related, the {\it too-big-to-fail} (TBTF) problem states that a substantial population of massive concentrated subhaloes produced in DMO simulations are missing in the observed satellite or field dwarfs in the Local Group~\citep{MBK2011,MBK2012,Tollerud2014,SGK2014,Kirby2014} and beyond~\citep{Papastergis2015}. 

Although the inclusion of stellar/supernovae feedback process within the $\Lambda {\rm CDM}$ framework has been shown to alleviate many of these tensions~\citep[e.g.,][]{Governato2010,Pontzen2012,Madau2014,Brooks2014,Wetzel2016,Sawala2016,SGK2019}, a population of compact dwarfs is often absent in CDM simulations (plus baryons) that produce diffuse dark matter cores~\citep[e.g.,][]{Santos2018,Jiang2019,SGK2019}. Related to this point, some have argued that the rotation curves (and hence the mass distribution) of dwarf galaxies may be more diverse than CDM predictions in the field~\citep{Oman2015} and Milky Way satellites~\citep{Kaplinghat2019} (recently revisited in \citealt{Santos2018} that baryon-induced core formation predictions do not obviously agree with the correlation between the degree of coring and baryon dominance). Therefore, it is important to explore how non-standard dark matter models (in conjunction with baryonic physics) behave at sub-galactic scales, and whether they can address these small-scale problems coherently. 

Self-interacting dark matter (SIDM), as an important category of alternative dark matter models, has been proposed and discussed in the literature for about three decades~\citep[e.g.,][]{Carlson1992, deLaix1995, Firmani2000,Spergel2000}. SIDM is well motivated by hidden dark sectors as extensions to the Standard Model~\citep[e.g.,][]{Ackerman2009,Arkani-Hamed2009,Feng2009,Feng2010,Loeb2011,vandenAarssen2012,CyrRacine2013,Tulin2013,Cline2014}. And the introduction of SIDM could potentially address some small-scale problems~\citep[see the review of][and references therein]{Tulin2018} through the thermalization processes of dark matter at galaxy centers. Previous DMO SIDM simulations have found that a self-interaction cross-section of $\sim 1\cpm$ could address the {\it core-cusp} and TBTF problems in dwarf galaxies~\citep[e.g.,][]{Vogelsberger2012, Rocha2013, Zavala2013, Elbert2015}. In addition, SIDM models with comparable cross-sections also have the potential to influence~\citep[e.g.,][]{Kamada2017,Creasey2017,Kahloefer2019,Sameie2020} the diversity of rotation curves of field dwarfs~\citep{Oman2015} and the diverse central densities of Local Group satellites~\citep{Kaplinghat2019}. 

\begin{table}
    \centering
    \begin{tabular}{p{0.06\textwidth}|p{0.05\textwidth}|p{0.05\textwidth}|p{0.05\textwidth}|p{0.05\textwidth}|p{0.06\textwidth}}
        \hline
        Simulation & $M^{\rm cdm}_{\rm halo}$ & $R^{\rm cdm}_{\rm vir}$ & $M^{\rm cdm}_{\ast}$ & $r^{\rm cdm}_{1/2}$ & $r^{\rm conv}_{\rm dm}$\\
        name & $[\msun]$ & $[\kpc]$ & $[\msun]$ & $[\kpc]$ & $[\pc]$ \\
        \hline
        \hline
    \end{tabular}
    
    \textbf{Ultra faint dwarf\\}
    \begin{tabular}{p{0.06\textwidth}|p{0.05\textwidth}|p{0.05\textwidth}|p{0.05\textwidth}|p{0.05\textwidth}|p{0.06\textwidth}}
        m09	  & 2.5e9 & 35.6 & 7.0e4 & 0.46 & 65   \\
        \hline
    \end{tabular}
    
    \textbf{Classical dwarfs\\}
    \begin{tabular}{p{0.06\textwidth}|p{0.05\textwidth}|p{0.05\textwidth}|p{0.05\textwidth}|p{0.05\textwidth}|p{0.06\textwidth}}
        m10b& 9.4e9 & 55.2 & 5.8e5 & 0.36 & 77 \\
        m10q& 7.5e9 & 51.1 & 1.7e6 & 0.72 & 73 \\
        m10v& 8.5e9 & 53.5 & 1.4e5 & 0.32 & 65 \\
        \hline
    \end{tabular}
    
    \textbf{Bright dwarfs\\}
    \begin{tabular}{p{0.06\textwidth}|p{0.05\textwidth}|p{0.05\textwidth}|p{0.05\textwidth}|p{0.05\textwidth}|p{0.06\textwidth}}
        m11a & 3.6e10 & 86.7 & 3.7e7 & 1.2 & 310 \\
        m11b & 4.2e10 & 90.7 & 4.2e7 & 1.7 & 250 \\
        m11q & 1.5e11 & 138.7 & 2.9e8 & 3.1 & 240 \\
        \hline
    \end{tabular}

   \textbf{ Milky Way-mass hosts\\}
    \begin{tabular}{p{0.065\textwidth}|p{0.05\textwidth}|p{0.05\textwidth}|p{0.05\textwidth}|p{0.05\textwidth}|p{0.06\textwidth}}
        m11f      & 4.5e11 & 200.2 & 1.0e10 & 2.9 & 280 \\
        m12i l.r. & 1.1e12 & 272.3 & 1.1e11 & 2.0 & 290 \\
        m12f l.r. & 1.5e12 & 302.8 & 1.3e11 & 4.1 & 310 \\
        m12m l.r. & 1.5e12 & 299.3 & 1.4e11 & 6.1 & 360 \\
        m12i h.r. & 9.8e11 & 259.9 & 2.4e10 & 3.7 & 150 \\
        \hline
    \end{tabular}
    
    \caption{ \textbf{Simulations of the FIRE-2 dSIDM suite.} The simulated galaxies are labelled and grouped by their halo masses. They are classified into four categories: ultra faint dwarfs; classical dwarfs, with typical halo mass $\lesssim 10^{10} \msun$; bright dwarfs, with typical halo mass $\sim 10^{10-11} \msun$; Milky Way-mass galaxies, with typical halo mass $\sim 10^{12}\msun$. These haloes are randomly picked from the standard FIRE-2 simulation suite~\citep{Hopkins2018}, sampling various star formation and merger histories. All units are physical. \newline \hspace{\textwidth}
    (\textbf{1}) Name of the simulation. ``l.r.'' (``h.r.'') indicates low (high)-resolution version of the simulation. Typically, the high-resolution version has $8$ times better mass resolution. \newline \hspace{\textwidth}
    (\textbf{2}) $M^{\rm cdm}_{\rm halo}$, $R^{\rm cdm}_{\rm vir}$: Virial mass and virial radius of the halo (defined using the overdensity criterion introduced in \citealt{Bryan1998}) in the CDM simulation with baryons at $z=0$. \newline \hspace{\textwidth}
    (\textbf{3}) $M^{\rm cdm}_{\ast}$, $r^{\rm cdm}_{\rm 1/2}$: Galaxy stellar mass and galaxy stellar half mass radius (defined using stellar particles within 10\% $R_{\rm vir}$) in the CDM simulation at $z=0$. \newline \hspace{\textwidth}
    (\textbf{4}) $r^{\rm conv}_{\rm dm}$: Radius of convergence in dark matter properties at $z=0$ (calculated for the CDM DMO simulations in the standard FIRE-2 series based on the \citealt{Power2003} criterion). As shown in \citet{Hopkins2018}, the convergence radii in simulations with baryons can in fact extend to much smaller radii. \newline \hspace{\textwidth}
    }
    \label{tab:sim}
\end{table}

The astrophysical studies of SIDM mentioned above all focused on {\em elastic} dark matter self-interactions. However, in most generic particle physics realizations of SIDM, dark matter would have inelastic (or specifically dissipative) self-interactions~\citep[e.g.,][]{Arkani-Hamed2009,Alves2010,Kaplan2010,Loeb2011,CyrRacine2013,Cline2014,Boddy2014,Wise2014,Foot2015b,Schutz2015,Boddy2016,Finkbeiner2016,Zhang2017,Blennow2017,Gresham2018}. The impact of dissipative processes of dark matter is rarely explored in the context of cosmological structure formation. Indeed, much of the focus of previous studies of purely elastic SIDM (eSIDM, references above) was on how it might influence some of the small-scale questions above by kinematically heating dark matter and lowering central halo densities. However, other processes can strongly influence these properties. For instance, gas outflows driven by stellar/supernovae feedback could help generate dark matter cores~\citep{Governato2010,Governato2012,Pontzen2012,Madau2014}, a process which has been verified in CDM simulations with baryons~\citep[e.g.,][]{Brooks2014,Dutton2016,Fattahi2016,Sawala2016,Wetzel2016,SGK2019,Buck2019}. In subsequent eSIDM simulations with baryons~\citep[e.g.,][]{Vogelsberger2014,Elbert2015,Fry2015,Robles2017,Despali2019,Fitts2019,Robles2019}, the distinct signatures of eSIDM were found to be substantially reduced with the inclusion of baryons. In principle, this could hide dark matter physics that would otherwise lead to {\em enhanced} central densities in DMO simulations. The parameter space for dSIDM, as an example of such models, is therefore much more widely open than previously believed, due to these recent developments. Beyond that, the halo contraction driven by dissipative interactions could potentially produce compact dwarf and/or promote the diversity of dwarf galaxy rotation curves to the observed level.

\begin{figure*}
    \centering
    %\textbf{ Classical dwarfs\\} \vspace{0.1cm}
    \includegraphics[width=0.245\textwidth]{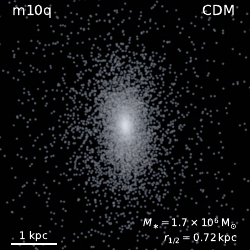}
    \includegraphics[width=0.245\textwidth]{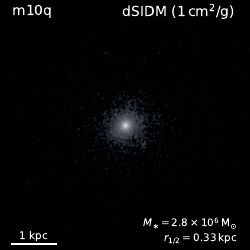}
    \includegraphics[width=0.245\textwidth]{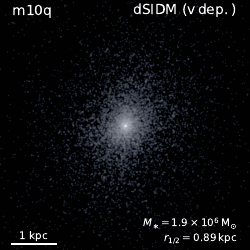}
    \includegraphics[width=0.245\textwidth]{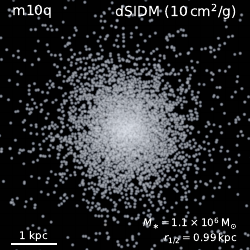}
    
    \includegraphics[width=0.245\textwidth]{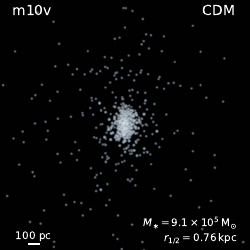}
    \includegraphics[width=0.245\textwidth]{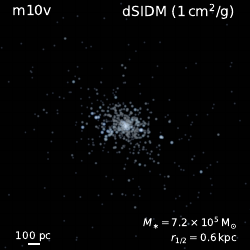}
    \includegraphics[width=0.245\textwidth]{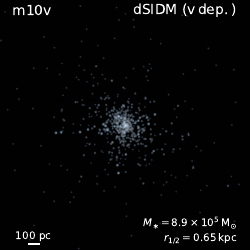}
    \includegraphics[width=0.245\textwidth]{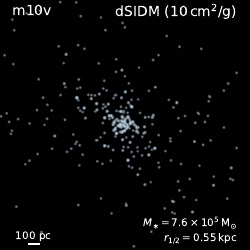}
    
    \caption{\textbf{Visualizations of two simulated classical dwarfs.} Each column corresponds to one dark matter model studied. The images are mock {\it Hubble Space Telescope} composites of u,g,r bands with a logarithmic surface brightness stretch. We use the {\sc STARBURST99} model to determine the SED of each stellar particle based on its age and initial metallicity and use ray-tracing~\citep{Hopkins2005} to model dust attenuation assuming a Milky Way-like reddening curve and a dust-to-metal ratio of $0.4$. The side lengths of the images are chosen to be $8\times r_{\rm 1/2}$ of the CDM run. The dSIDM models with $(\sigma/m)=1\cpm$ and the velocity-dependent cross-section produce visibly more concentrated stellar content compared to the CDM case (the effective cross-section as defined in \citetalias{Shen2021} of our velocity-dependent model in classical dwarfs is about $0.3\cpm$). However, the model with $(\sigma/m)=10\cpm$ produces overall fluffier stellar distribution.
    }
    \label{fig:image1}
\end{figure*}

\begin{figure*}
    \centering
    %\textbf{ Bright dwarfs\\}\vspace{0.1cm}
    \includegraphics[width=0.245\textwidth]{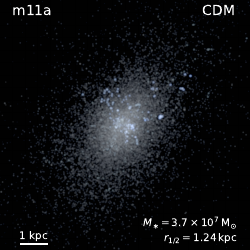}
    \includegraphics[width=0.245\textwidth]{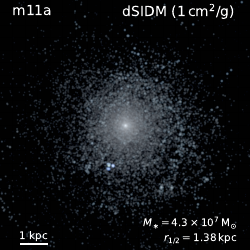}
    \includegraphics[width=0.245\textwidth]{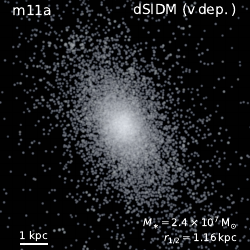}
    \includegraphics[width=0.245\textwidth]{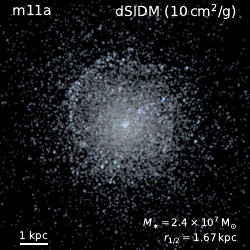}
    
    \includegraphics[trim=0 0.7cm 0 0.7cm, clip, width=0.245\textwidth]{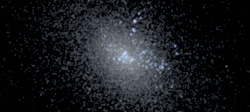}
    \includegraphics[trim=0 0.7cm 0 0.7cm, clip, width=0.245\textwidth]{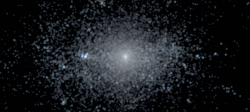}
    \includegraphics[trim=0 0.7cm 0 0.7cm, clip, width=0.245\textwidth]{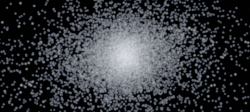}
    \includegraphics[trim=0 0.7cm 0 0.7cm, clip, width=0.245\textwidth]{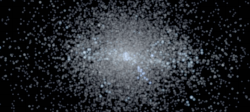}
    
    \includegraphics[width=0.245\textwidth]{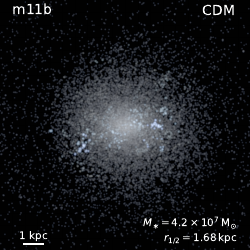}
    \includegraphics[width=0.245\textwidth]{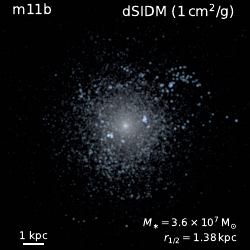}
    \includegraphics[width=0.245\textwidth]{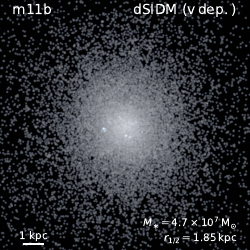}
    \includegraphics[width=0.245\textwidth]{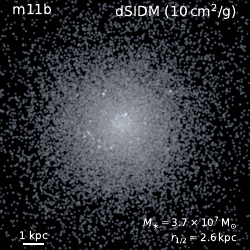}
    
    \includegraphics[trim=0 0.7cm 0 0.7cm, clip, width=0.245\textwidth]{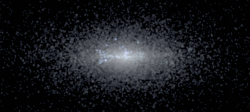}
    \includegraphics[trim=0 0.7cm 0 0.7cm, clip, width=0.245\textwidth]{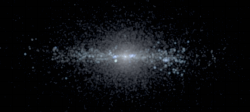}
    \includegraphics[trim=0 0.7cm 0 0.7cm, clip, width=0.245\textwidth]{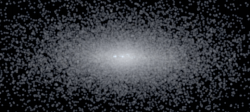}
    \includegraphics[trim=0 0.7cm 0 0.7cm, clip, width=0.245\textwidth]{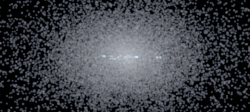}
    
    \includegraphics[width=0.245\textwidth]{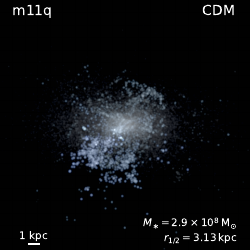}
    \includegraphics[width=0.245\textwidth]{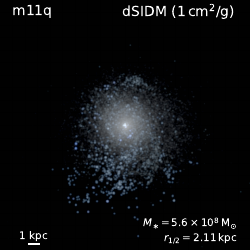}
    \includegraphics[width=0.245\textwidth]{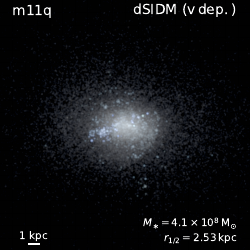}
    \hspace{0.245\textwidth}

    \includegraphics[trim=0 0.7cm 0 0.7cm, clip, width=0.245\textwidth]{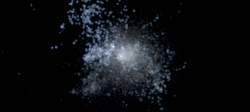}
    \includegraphics[trim=0 0.7cm 0 0.7cm, clip, width=0.245\textwidth]{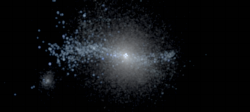}
    \includegraphics[trim=0 0.7cm 0 0.7cm, clip, width=0.245\textwidth]{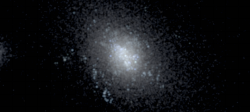}
    \hspace{0.245\textwidth}
    
    \caption{\textbf{Visualizations of three simulated bright dwarfs.} The images are generated in the same way as those in Figure~\ref{fig:image1}. Since some of the bright dwarfs develop disk-like structures, we show both face-on and edge-on images here. Compared to the CDM case, the stellar disks in the dSIDM model with $(\sigma/m)=1\cpm$ are more well-defined and exhibit more concentrated central regions. On the other hand, the velocity-dependent dSIDM model produces galaxies that are visibly similar to the CDM case given its small effective cross-section at this mass scale ($(\sigma_{\rm eff}/m) \sim 0.01\cpm$). Interestingly, the model with $(\sigma/m)=10\cpm$ produces stellar disks accompanied by overall fluffier stellar distribution compared to the model with $(\sigma/m)=1\cpm$ and CDM.
    }
    \label{fig:image2}
\end{figure*}

Meanwhile, continuous improvements in observations of local dwarf galaxies and other small-scale baryonic structures have enabled great opportunities to constrain the nature of dark matter. For example, the census of ultra-faint satellite galaxies in the Local Group through optical imaging surveys has been boosted in recent years, using the data from the Dark Energy Survey \citep[DES;][]{DES2016,Drlica2015,Bechtol2015,Drlica2020}, the Panoramic Survey Telescope and Rapid Response System \citep[Pan-STARRS;][]{Laevens2015, Laevens2015b}, and others \citep[e.g.,][]{Torrealba2016, Torrealba2019}. Many of the recently detected ultra-faints appear to be clustered around the Large Magellanic Cloud (LMC; \citealt{Drlica2015,Koposov2015}). These candidate LMC satellites are attractive targets for ongoing and future observations to test the $\Lambda {\rm CDM}$ model~\citep{Wheeler2015}. The structural and dynamical properties of the Local Group satellites with resolved stellar populations have been measured~\citep[see for example compilations by][and references therein]{McConnachie2012,Tollerud2014} and play a key role in understanding the TBTF problem~\citep{MBK2011}. In the near future, the Legacy Survey of Space and Time \citep[LSST,][]{LSST2009} at the Vera Rubin Observatory has the potential to substantially expand the discovery space of faint dwarf galaxies, being sensitive to galaxies one hundred times fainter than Sloan Digital Sky Survey \citep[SDSS,][]{York2000} at the same distance~\citep{Bullock2017}. Beyond the Local Group, the Dark Energy Camera \citep[DECam,][]{DECam2015} and Subaru (Hyper) Suprime-Cam~\citep[e.g.,][]{Miyazaki2002,Miyazaki2018} are being used to search for faint companions of nearby galaxies~\citep[e.g.,][]{Sand2015,Crnojevic2016,Carlin2016}, as well as the LSBs and ultra-diffuse dwarf galaxies (UDGs) in cluster environment~\citep[e.g.,][]{Koda2015,Munoz2015,Mihos2015,VanDokkum2015,Eigenthaler2018}. In addition, for relatively massive disky dwarfs (late-type), radio observations have reported the H\Rmnum{1} rotation curves and mass models of a few hundred of them in the Local Universe~\citep[e.g.,][]{Oh2011,Ott2012,Oh2015,Lelli2016}. The time is therefore ripe to make testable predictions from different dark matter model parameter space.

In \citetalias{Shen2021} of this series, we have introduced the first suite of cosmological hydrodynamical zoom-in simulations of galaxies in dSIDM. The simulations employ the FIRE-2 model~\citep{Hopkins2018} for hydrodynamics and galaxy formation physics and the dSIDM model is parameterized by a dimensionless dissipation factor and a constant (or velocity-dependent) self-interaction cross-section. We explored the parameter choices that lead to weakly-collisional and weakly-dissipative dark fluid on cosmological scale. Similar dSIDM models have been explored using semi-analytical methods and idealized simulations~\citep{Essig2019,Huo2019}. We stress that this type of model is qualitatively different from the dissipative dark matter proposed by many previous studies~\citep{Fan2013,Foot2013,Randall2015}, which are highly-coupled (effective ``collision'' rate, $\Gamma_{\rm eff} \gg H$, $H$ is the Hubble parameter) and highly-dissipative (similar to the cooling in the baryonic sector) and will likely give rise to disk-like structures or fragmentation to ``compact'' objects. In \citetalias{Shen2021}, we have identified several distinct properties of dwarf galaxies in dSIDM, including the steepening of the central density profile, enhanced rotation support of dark matter and the deformation of haloes to become oblate in shape. These unique structural and kinematic properties are expected to give observational signatures that could be used to constrain dSIDM models.

In this paper, we make predictions for various observed properties of galaxies in dSIDM and compare them to the observed dwarf satellite galaxies in the Local Group or LSBs in the field. The paper will be organized as follows: In Section~\ref{sec:sim}, we briefly review the simulation setup and the dark matter models studied. In Section~\ref{sec:baryonic-content}, we present predictions for the stellar content of the simulated dwarfs, including density profiles, size and mock optical images. Then in Section~\ref{sec:rotcurve}, the circular velocity profiles of the simulated dwarfs (or satellites of Milky Way-mass hosts) will be compared with their observational counterparts specifically. In Section~\ref{sec:satellite}, the satellite counts of simulated Milky Way-mass hosts will be studied. Finally, in Section~\ref{sec:conclusion}, the summary and conclusions will be presented.

\begin{figure*}
    \centering
    \includegraphics[width=0.245\textwidth]{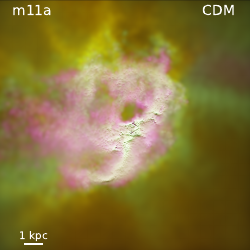}
    \includegraphics[width=0.245\textwidth]{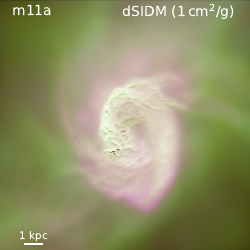}
    \includegraphics[width=0.245\textwidth]{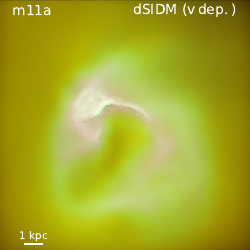}
    \includegraphics[width=0.245\textwidth]{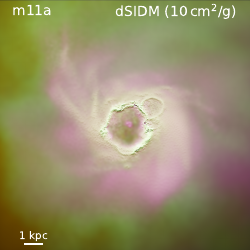}
    \includegraphics[trim=0 0.7cm 0 0.7cm, clip, width=0.245\textwidth]{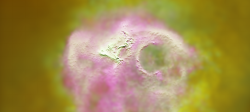}
    \includegraphics[trim=0 0.7cm 0 0.7cm, clip, width=0.245\textwidth]{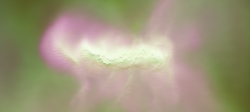}
    \includegraphics[trim=0 0.7cm 0 0.7cm, clip, width=0.245\textwidth]{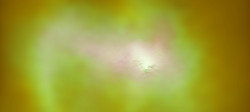}
    \includegraphics[trim=0 0.7cm 0 0.7cm, clip, width=0.245\textwidth]{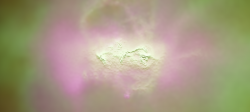}
    
    \includegraphics[width=0.245\textwidth]{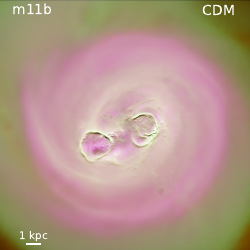}
    \includegraphics[width=0.245\textwidth]{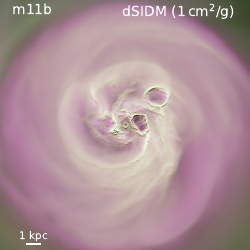}
    \includegraphics[width=0.245\textwidth]{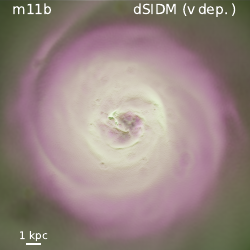}
    \includegraphics[width=0.245\textwidth]{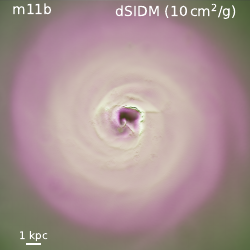}
    \includegraphics[trim=0 0.7cm 0 0.7cm, clip, width=0.245\textwidth]{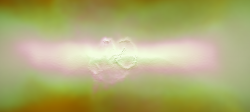}
    \includegraphics[trim=0 0.7cm 0 0.7cm, clip, width=0.245\textwidth]{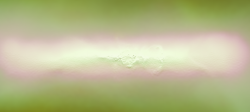}
    \includegraphics[trim=0 0.7cm 0 0.7cm, clip, width=0.245\textwidth]{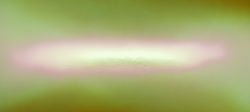}
    \includegraphics[trim=0 0.7cm 0 0.7cm, clip, width=0.245\textwidth]{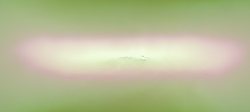}
    
    \includegraphics[width=0.245\textwidth]{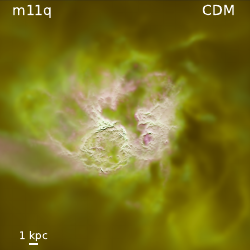}
    \includegraphics[width=0.245\textwidth]{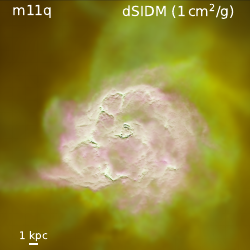}
    \includegraphics[width=0.245\textwidth]{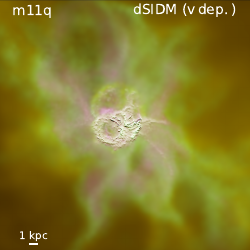}
    \hspace{0.245\textwidth}
    
    \includegraphics[trim=0 0.7cm 0 0.7cm, clip, width=0.245\textwidth]{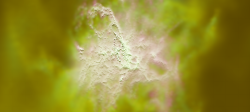}
    \includegraphics[trim=0 0.7cm 0 0.7cm, clip, width=0.245\textwidth]{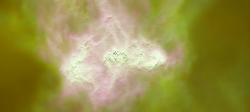}
    \includegraphics[trim=0 0.7cm 0 0.7cm, clip, width=0.245\textwidth]{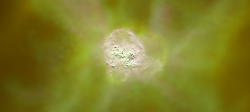}
    \hspace{0.245\textwidth}
    
    \caption{\textbf{Visualization of the gas content of three simulated bright dwarfs.} The images are logarithmically-weighted gas surface density projections. Each column corresponds to one dark matter model studied and each row corresponds to one bright dwarf simulated. For each dwarf, both the face-on and edge-on images are shown. The side lengths of the images are chosen as $12\times r^{\rm cdm}_{1/2}$. Each image is a composite of gas distribution in three phases characterized by the gas temperature. The magenta color represents the ``cold'' neutral gas with $T \lesssim 8000\,{\rm K}$; the green color represents the ``warm'' gas with $T \sim 1 \operatorname{-} 3\times 10^{4}\,{\rm K}$; the red color represents the ``hot'' ionized gas in the CGM with $T \gtrsim 10^{5}\,{\rm K}$. The neutral gas disks are promoted in the dSIDM-c1 and c10 models, even in m11a which is strongly perturbed by supernovae feedback in CDM.}
    \label{fig:gas_image}
\end{figure*}

\section{Simulations}
\label{sec:sim}

\subsection{Overview of the simulation suite}

The analysis in this paper is based on the FIRE-2 dSIDM simulation suite introduced in \citetalias{Shen2021}, which consists of $\sim\,$40 cosmological hydrodynamical zoom-in simulations of galaxies chosen at representative mass scales with CDM, eSIDM and dSIDM models. These simulations are part of the Feedback In Realistic Environments project~\citep[FIRE, ][]{Hopkins2014}, specifically the ``FIRE-2'' version of the code with details described in \citet{Hopkins2018}. The simulations adopt the code {\sc Gizmo}~\citep{Hopkins2015}, with hydrodynamics solved using the mesh-free Lagrangian Godunov ``MFM'' method. The simulations include cooling and heating from a meta-galactic radiation background and stellar sources in the galaxies, star formation in self-gravitating molecular gas and stellar/supernovae/radiation feedback. The FIRE physics, source code, and numerical parameters are identical to those described in \citet{Hopkins2018,SGK2019b} \footnote{We note that the CDM runs are rerun to exact match the configuration of dSIDM runs, so galaxy properties are not expected to be ``identical'' to the original FIRE-2 results.}. The minimum baryonic particle masses reached in these simulations are $m_{\rm b} \simeq 250\, (7000)\msun$ for dwarf (Milky Way-mass) galaxies. Further details of the numerical implementation are described in \citet{Hopkins2018} and \citetalias{Shen2021}. A full list of the galaxies simulated and relevant parameters are shown in the Table~\ref{tab:sim}. 

Dark matter self-interactions are simulated in a Monte-Carlo fashion following the implementation in \citet{Rocha2013}. In this paper, we study a simplified empirical dSIDM model: two dark matter particles lose a constant fraction (the dissipation factor $f_{\rm diss}$) of their kinetic energy in the center of momentum frame when they collide with each other. The extreme version of this type of interaction is the fusion process (i.e. $f_{\rm diss}=1$) of dark matter composites discussed in the context of self-interacting asymmetric dark matter \citep[e.g.,][]{Wise2014,Wise2015,Detmold2014,Krnjaic2015,Gresham2018} and specifically the dark ``nuggets'' model \citep{Wise2014,Gresham2018}. It is worth noting that there are other particle physics models for dSIDM~\citep[e.g.,][]{Alves2010,Kaplan2010,Fan2013,Boddy2014,Cline2014,Schutz2015,Foot2015} with potentially different behaviors on cosmological scale that are not captured by this simplified parameterization. However, it is a reasonable starting point to study the phenomenology of dissipative dark matter in cosmic structural formation.

The simulations employed a fiducial dissipation factor $f_{\rm diss} = 0.5$ and we explore models with constant self-interaction cross-section $(\sigma/m) = 0.1,1,10\,\cpm$ or a velocity-dependent cross-section model
\begin{equation}
    \dfrac{\sigma(v)}{m} = \dfrac{(\sigma/m)_{0}}{1+(v/v_0)^4},
    \label{eq:cx_vel}
\end{equation}
where the fiducial choice is $(\sigma/m)_{0} = 10 \cpm$ and $v_0 = 10 \kms$. The velocity dependence of the self-interaction cross-section is empirically motivated by the relatively tight constraints on SIDM at galaxy cluster scale~\citep[e.g.,][]{Markevitch2004,Randall2008,Kaplinghat2016} and the relatively high cross-section needed to strongly influence some small-scale galaxy properties~\citep[e.g.,][]{Vogelsberger2012,Rocha2013,Zavala2013,Elbert2015,Kaplinghat2016}. The velocity dependence is also a generic feature of many particle physics models for SIDM~\citep[e.g.,][]{Feng2009,Kaplan2010,CyrRacine2013, Cline2014,Boddy2014,Boddy2016,Zhang2017, Tulin2018}. Given the choices of cross-section and dissipation factor, the typical collision and energy dissipation time scale of dark matter will be smaller than the Hubble time scale but still larger than the free-fall time scale (see \citetalias{Shen2021} for details). In this regime, the dissipative dark matter is weakly-collisional and weakly-dissipative compared to the baryonic gas and will not fragment or form ``compact'' dark objects. In this paper, we will refer to the dSIDM model with constant cross-section $(\sigma/m) = 0.1,1,10\,\cpm$ as ``dSIDM-c0.1,1,10'' for simplicity. For comparison, a subset of the galaxies in the suite are also simulated with the eSIDM model with constant cross-section $(\sigma/m) = 1\,\cpm$.

\subsection{Host halo and substructures}
\label{sec:sim_substructure}

The simulations in this suite are all identified with the main ``target'' halo around which the high-resolution zoom-in region is centered. The central position and velocity of this main halo are defined by the center of mass (of dark matter particles) and are calculated via an iterative zoom-in approach. However, specifically for the measurements of stellar density profiles and galaxy visible sizes, we use stellar particles for the center identification. The bulk properties of the halo and the galaxy it hosts are calculated following the practice of the standard FIRE-2 simulations, as described in \citet{Hopkins2018}. We define the halo mass $M_{\rm halo}$ and the halo virial radius $R_{\rm vir}$ using the redshift-dependent overdensity criterion in \citet{Bryan1998}. We define the stellar mass $M_{\ast}$ as the total mass of all the stellar particles within an aperture of $0.1\,R_{\rm vir}$ and correspondingly define the stellar half-mass radius $r_{1/2}$ as the radius that encloses half of the total stellar mass. 

To identify substructures in post-processing, we take two steps following \citet{Wetzel2016,Necib2019,SGK2019} and \citet{Samuel2020}. We first identify bound dark matter subhaloes (of the main ``target'' halo) using the {\sc Rockstar}~\citep{Behroozi2013} halo finder\footnote{The adapted version \citep{Wetzel2020} for {\sc Gizmo} is at \href{ https://bitbucket.org/awetzel/rockstar-galaxies/}{https://bitbucket.org/awetzel/rockstar-galaxies/}}, based only on dark matter particles. The force resolution of {\sc Rockstar} is conservatively set to be the same as the softening length of dark matter particles in simulations. To exclude misidentified subhaloes with a limited number of particles, we only keep subhaloes with mass $M_{\rm 200,m}>3\times 10^{6}\msun$\footnote{$M_{\rm 200,m}$ and $R_{\rm 200,m}$ are defined for the subhalo with the density criterion $200$ times the mean matter density of the Universe at $z=0$ calculated by {\sc Rockstar}. Note that this is different from the virial mass definition of the main ``target'' halo, and is used only for selection purpose.} and maximum circular velocity $V^{\rm max}_{\rm circ}>5\kms$ from the output {\sc Rockstar} halo catalogs. In the second step, stellar particles are assigned to the identified dark subhaloes through an iterative method~\citep{Wetzel2016,SGK2019,Wetzel2020,Samuel2020}. Initially, stellar particles are assigned to a dark matter subhalo with a generous cut on their distances to the subhalo center ($r\leq 0.8 R_{\rm 200,m}$ and $r\leq 30\kpc$) and velocities with respect to the subhalo center ($v\leq 2 V^{\rm max}_{\rm circ}$ and $v\leq 2 \sigma^{\rm 3d}_{\rm v,dm}$). Subsequently, stellar particles are iteratively removed if $r<1.5\, r_{\rm 90}$, where $r_{\rm 90}$ is the radius than enclose $90\%$ of the stellar mass currently associated to the subhalo, or if $v<2 \sigma^{\rm 3d}_{\rm v,\ast}$, where $\sigma^{\rm 3d}_{\rm v,\ast}$ is the three-dimensional velocity dispersion of stars currently associated to the subhalo, until the number of stellar particles selected stabilizes at one percent level. Finally, we define $M_{\ast}$ of the subhalo as the mass sum of all the stellar particles that remain assigned to each galaxy in this way and correspondingly define $r_{\rm 1/2}$ as the radius within which the enclosed mass is $M_{\ast}/2$. The mass density profiles will be calculated in spherical shells centering on each subhalo, using all relevant types of particles in those shells. The circular velocity will be calculated based on the total mass enclosed by each shell.

\begin{figure*}
    \centering
    \includegraphics[width=0.49\textwidth]{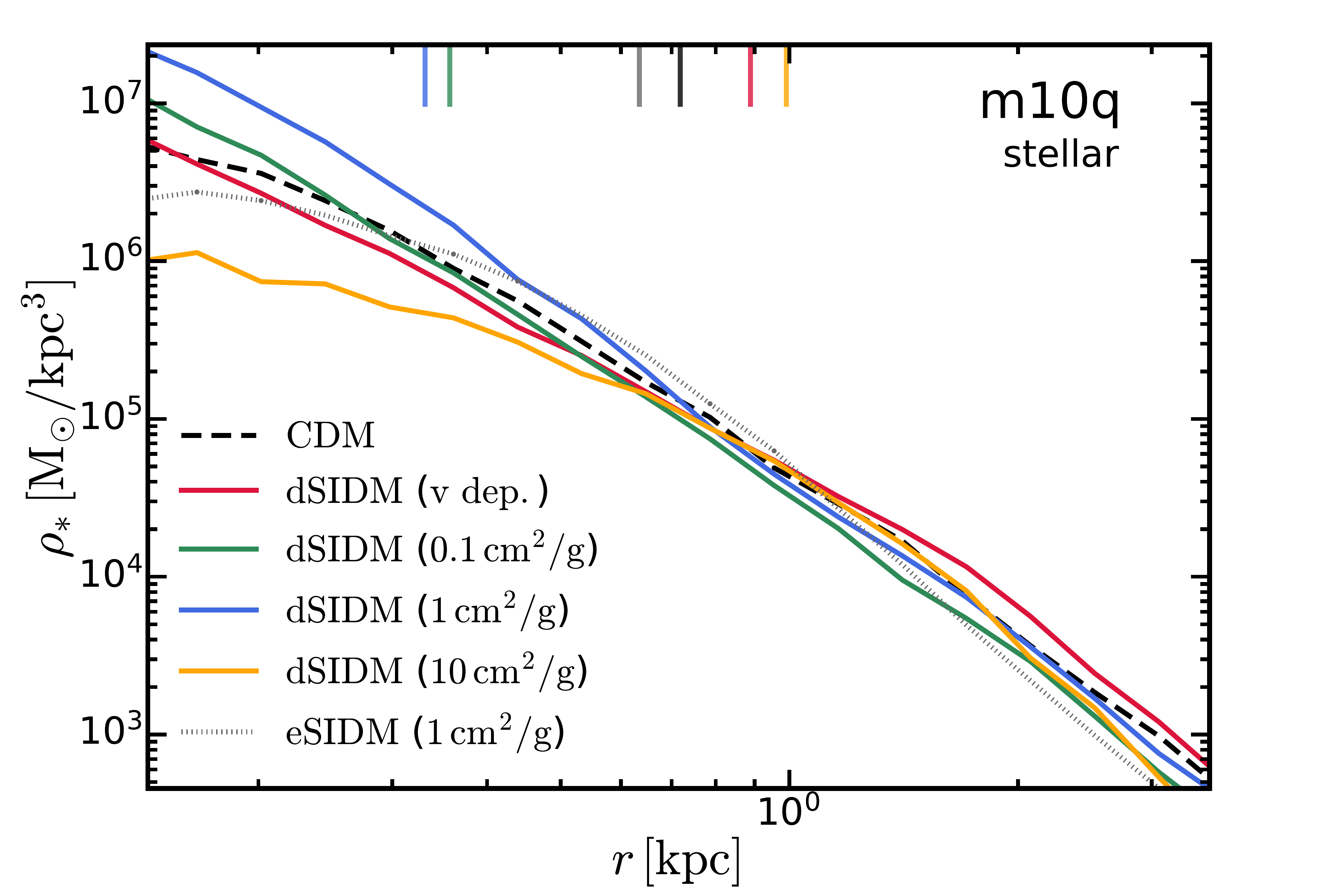}
    \includegraphics[width=0.49\textwidth]{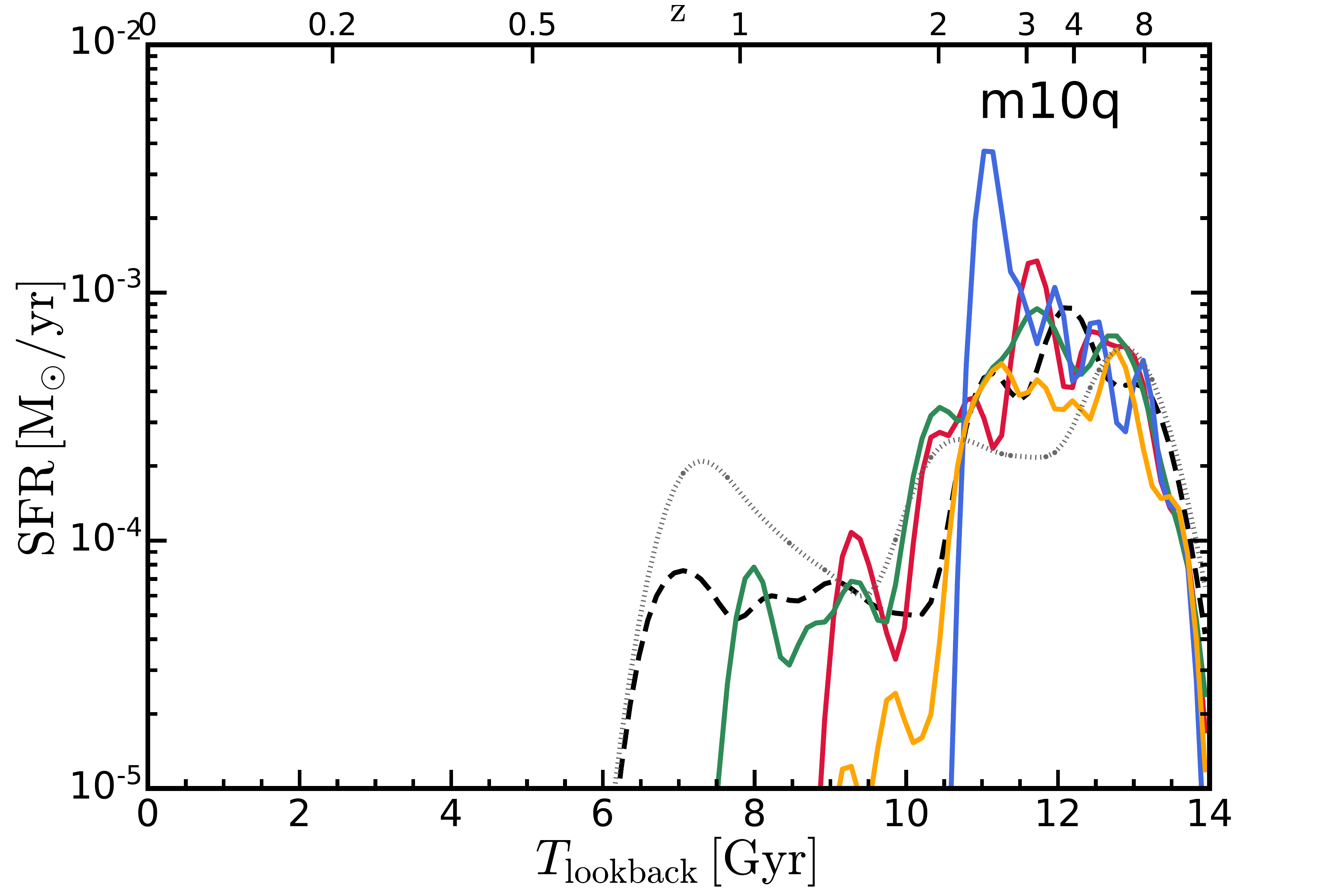}
    \includegraphics[width=0.49\textwidth]{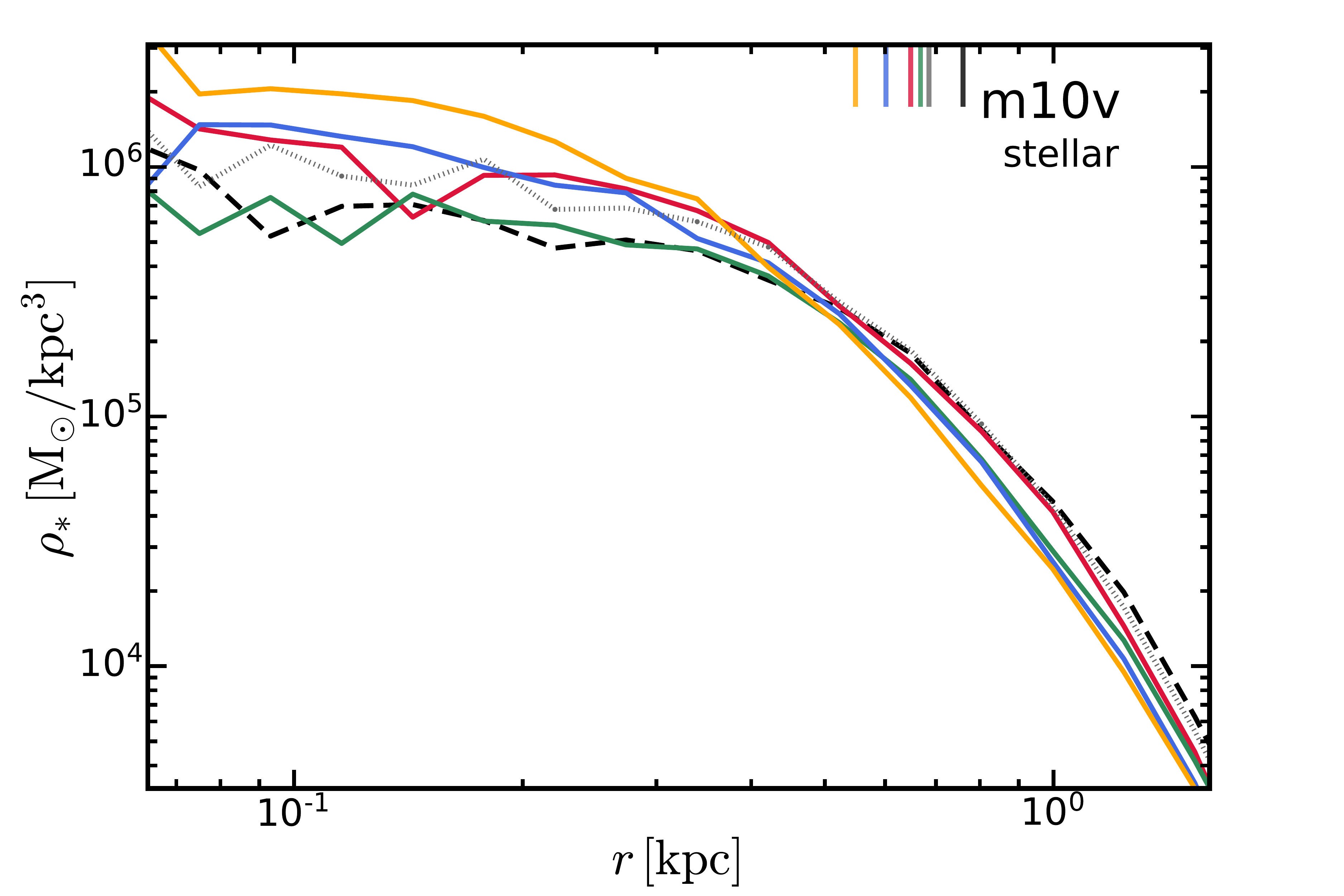}
    \includegraphics[width=0.49\textwidth]{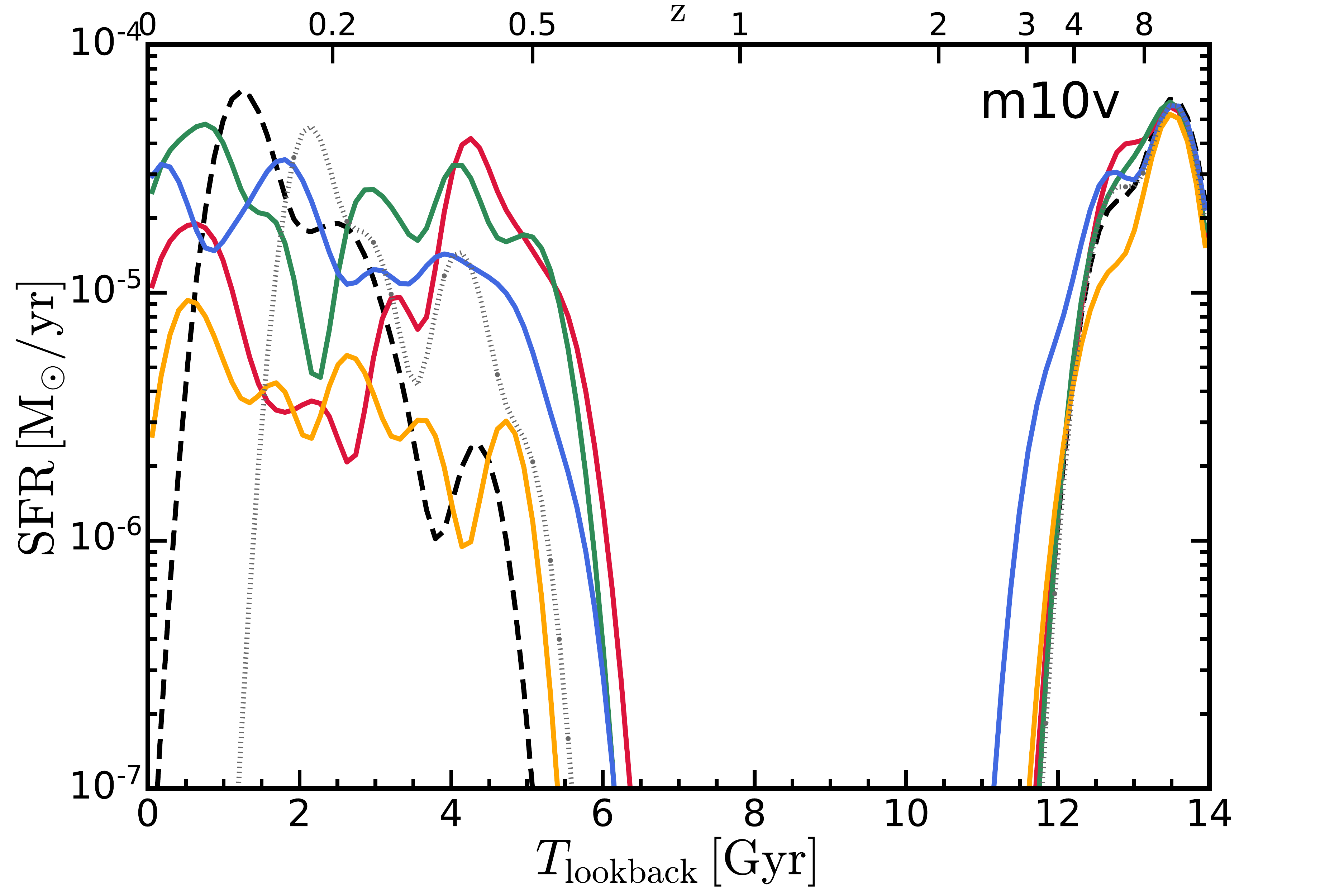}
    \caption{ {\it Left column}: \textbf{Stellar density profiles of simulated classical dwarfs.} The density profiles from different dark matter models are presented as labelled. The short vertical lines indicate the stellar-half-mass-radius of the galaxy in each model. The m10q and m10v haloes show different responses to dark matter dissipation. In m10q ({\em top}), which forms its stars early, a cuspy stellar profile appears with moderate dSIDM cross-sections accompanied by shrinking galaxy size, and then the profile turns shallower when the cross-section further increases. In m10v ({\em bottom}), which forms quite late, the profile becomes more concentrated monotonically as the cross-section increases, and the decline of galaxy size is less dramatic. This is related to the distinct star formation histories of the two galaxies as shown on the right. {\it Right column}: \textbf{Archaeological star formation history of simulated classical dwarfs.} This is computed as the age distribution of stellar particles within $10\%\,R^{\rm cdm}_{\rm vir}$ at $z=0$. The galaxy m10q has an early star formation history peaked at $z\simeq 3$. The stars have more time to react to the underlying dark matter distribution. On the other hand, the galaxy m10v with a relative late period of star formation does not exhibit this. The late time star formation and feedback also puffs up the stellar content and make it less dependent on the underlying dark matter distribution.}
    \label{fig:stellar-profile-classical}
\end{figure*}

\section{Galaxy baryonic content}
\label{sec:baryonic-content}

\subsection{Galaxy morphology}
\label{sec:baryonic-content-morphology}

In Figure~\ref{fig:image1} and Figure~\ref{fig:image2}, we show mock images of simulated dwarf galaxies at $z=0$, grouped as classical and bright dwarfs. Each image is a mock {\it Hubble Space Telescope} composite of u,g,r bands with a logarithmic stretch of the surface brightness. We use the STARBURST99 model~\citep{Leitherer1999} to determine the spectral energy distribution (SED) of each stellar particle based on its age and initial metallicity, and use the ray-tracing method~\citep{Hopkins2005} to model dust attenuation, assuming a Milky Way-like reddening curve and a dust-to-metal ratio of $0.4$. For the classical dwarfs (m10 galaxies) in Figure~\ref{fig:image1}, the dSIDM-c1 and the velocity-dependent model produce visibly more concentrated stellar content than the CDM case. The contraction of the stellar content is likely related to the contraction of the underlying dark matter distribution. On the other hand, the dSIDM-c10 model produces fluffier stellar content in m10q. This phenomenon is likely related to the lowered normalization of the central dark matter density profile, and thus shallower gravitational potential, in this model as found and described in in \citetalias{Shen2021}. However, in the same model, the stellar content of m10v is still compact, which demonstrates the large galaxy-to-galaxy variations of the star formation and corresponding dark matter and galaxy dynamics in classical dwarfs. This variation mainly comes from the distinct star formation histories of the two dwarfs. As will be shown in the following section (see also \citet{Hopkins2018}), m10q is an early-forming galaxy with half of its stellar mass formed at $z\gtrsim 4$ while m10v is late-forming with most of its stellar mass formed at $z\lesssim 0.4$, dominated by a few starburst events within the recent $4\Gyr$. Therefore, the early-formed stars in m10q would have enough time to relax and respond to the change of the underlying dark matter structure, while m10v is still strongly affected by its very recent star formation and feedback. Similar phenomena are found in the stellar density profiles of the two classical dwarfs presented in Figure~\ref{fig:stellar-profile-classical}, which will be discussed in Section~\ref{sec:baryonic-content-sdpro}.

\begin{figure*}
    \centering
    \includegraphics[width=0.49\textwidth]{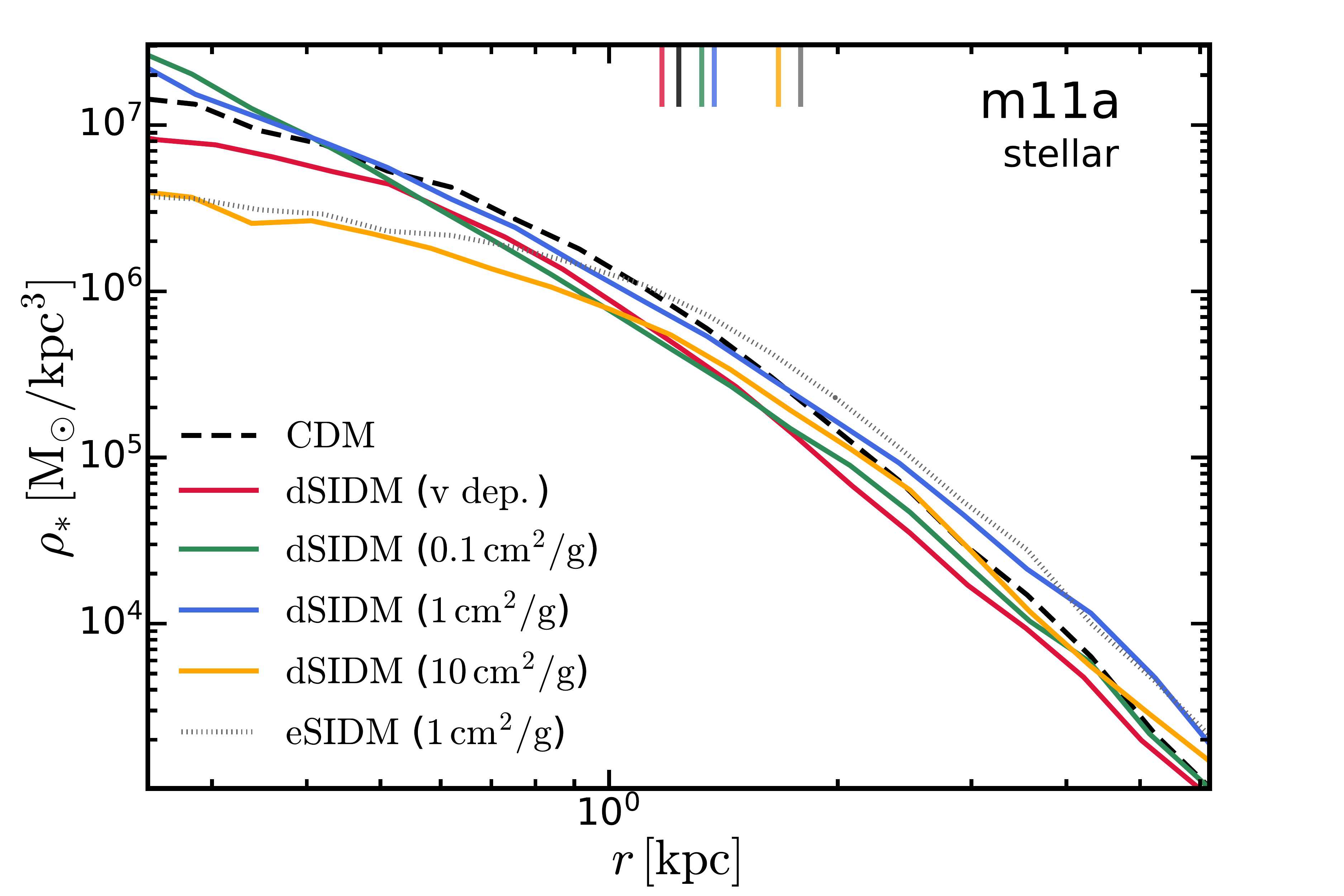}
    \includegraphics[width=0.49\textwidth]{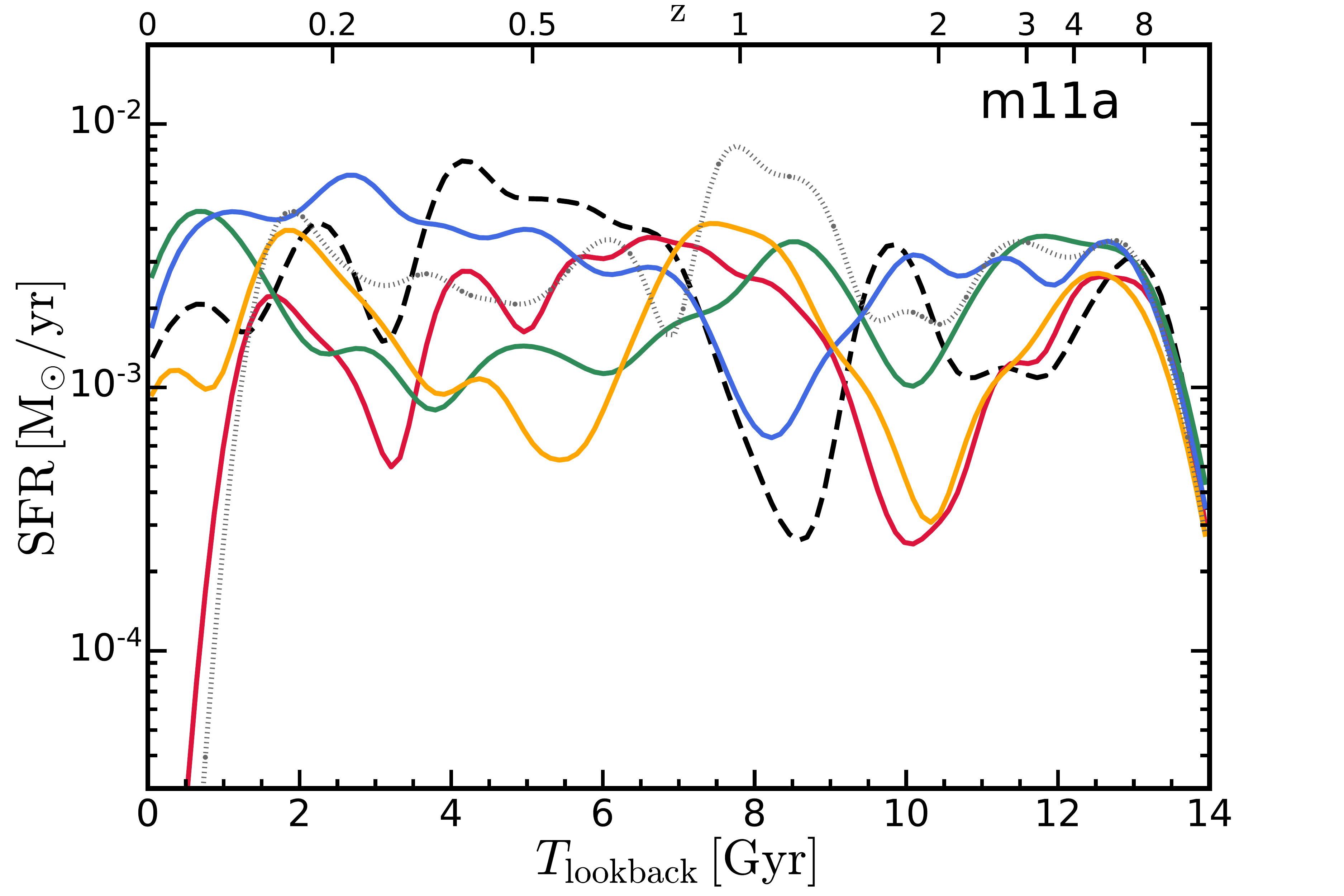}
    \includegraphics[width=0.49\textwidth]{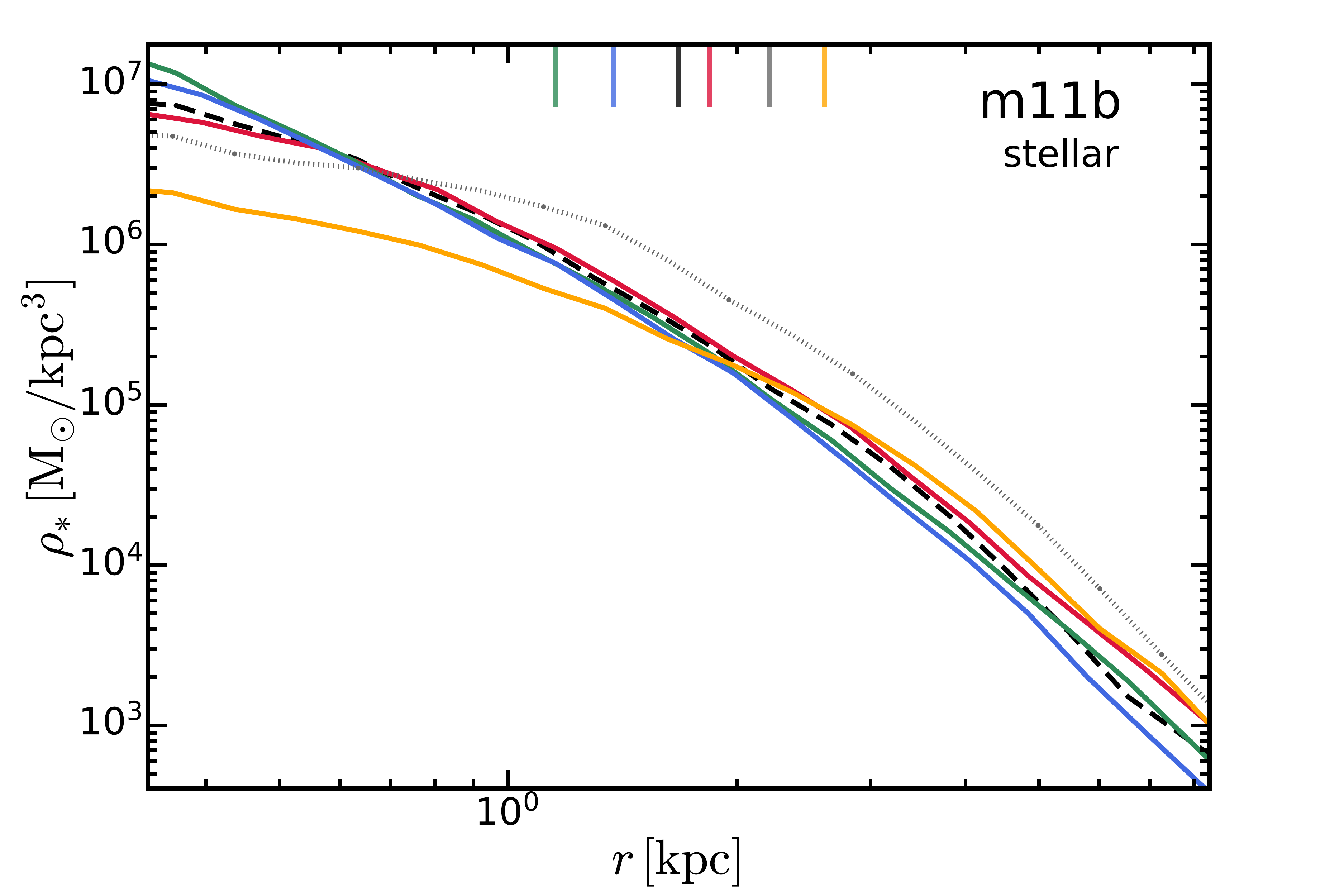}
    \includegraphics[width=0.49\textwidth]{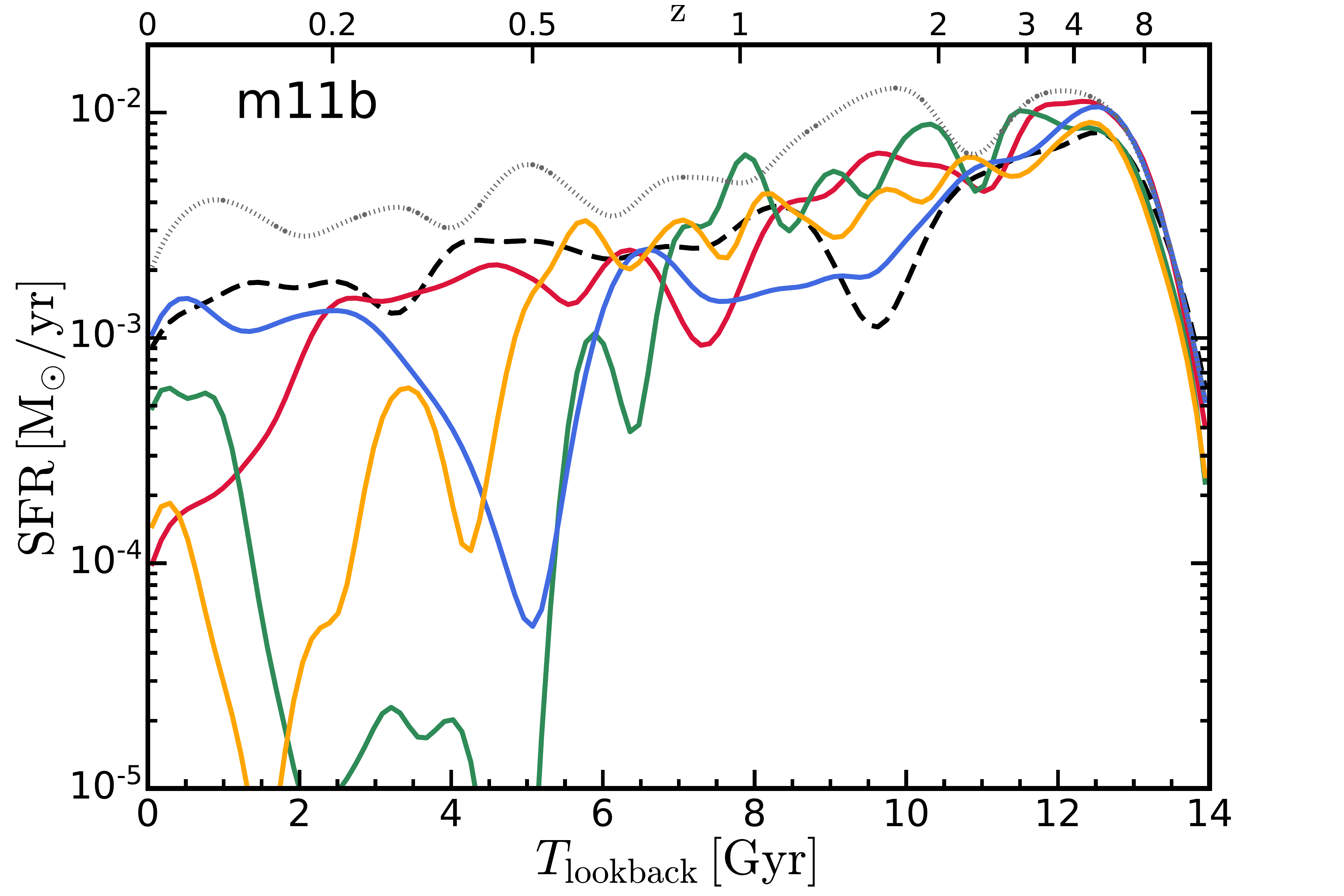}
    \caption{{\it Left column}: \textbf{Stellar density profiles of simulated bright dwarfs.} The notation is the same as Figure~\ref{fig:stellar-profile-classical}. In both m11a and m11b, the stellar density profiles become cuspy in dSIDM models with moderate cross-sections while turning shallower as we further increase the cross-section. This largely reflects similar behavior seen in the dark matter density profiles in \citetalias{Shen2021} -- in particular, at very high cross-sections the central dark matter profiles are flattened via dark rotation. {\it Right column}: \textbf{Archaeological star formation history of simulated bright dwarfs.} The notation is the same as Figure~\ref{fig:stellar-profile-classical}. The galaxy m11a has a relatively flat star formation history and is not significantly affected by the dark matter physics. However, in m11b, dips in star formation history at low redshifts appear in dSIDM models.}
    \label{fig:stellar-profile-bright}
\end{figure*}

For the bright dwarfs (m11 galaxies) in Figure~\ref{fig:image2}, we show both face-on and edge-on images because stellar disks start to show up in some simulated galaxies. The viewing angles are determined by the total angular momentum of the stellar particles with half of the field of view. Compared to its CDM counterpart, the dSIDM-c1 model gives rise to thinner and more well-defined stellar disks and meanwhile more concentrated central cusps. This again is in line with the more concentrated underlying dark matter distribution. The morphological transition is caused by the stronger central attraction forces provided by the compact cusps formed in dSIDM haloes. The central dSIDM cusp provides a well-defined ``center'' of the galaxy for star-forming gas to coherently rotate around and also stabilize the thin stellar disk formed. For the velocity-dependent model, the compactness of the stellar content is close to the CDM case because the effective cross-section (at the mass scale of bright dwarfs) decreases to $\sim 0.01\cpm$ which is much smaller than that in the classical dwarfs. Nevertheless, the stellar distribution in this model is more extended and the on-going star formation is also suppressed (see the lack of blue star-forming clouds in the images). For the dSIDM-c10 model, stellar disks are produced but accompanied by apparently fluffier stellar distributions, which is similar to what we described in classical dwarfs. 

An important feature of the bright dwarfs is the formation of co-rotating baryonic structures, e.g. the stellar disks in some of the m11 galaxies, which is absent in lower mass dwarfs. The larger halo mass and the presence of dense central baryonic components make these galaxies more stable against the energy/momentum injection from feedback, and therefore more able to sustain a rotationally supported gaseous disk~\citep[e.g.,][]{Obreja2016,ElBadry2018}. In observations, a highly-rotating subset of disky dwarf galaxies (late-type) have been found in H\Rmnum{1} surveys at similar mass scale~\citep[e.g.,][]{Oh2011,Oh2015,Lelli2016}. In Figure~\ref{fig:gas_image}, we show the gas surface density projections of the simulated dwarfs in the face-on and edge-on direction (determined by the angular momentum of the gas). The images are composites of the gas surface density in three phases, with the magenta/green/red color representing the ``cold'' neutral gas with $T \lesssim 8000\,{\rm K}$, the ``warm'' gas with $T \sim 1 \operatorname{-} 3\times 10^{4}\,{\rm K}$ and the ``hot'' ionized gas with $T \gtrsim 10^{5}\,{\rm K}$, respectively. The cold neutral gas in these dwarf galaxies is confined by the hot CGM gas, and star formation takes place in dense molecular clouds embedded in the gas disks, and perturbations from subsequent stellar/supernovae feedback manifest as ``super bubbles'' in the ISM. The feedback heats up the gas at the shock front of ``super bubbles'' and creates a warm layer in the gas disk. Among all three m11 galaxies simulated in CDM, only m11b develops a well-defined rotating disk consists of cold neutral gas while the other two dwarfs are severely perturbed by feedback. However, in dSIDM-c1, all three dwarfs show signatures of a co-rotating gaseous disk, with obvious diskness in the edge-on projection and spiral arms visible in the face-on projection. Similar to what we found for the stellar disk, the compact cusps of dSIDM haloes provide stronger central attraction forces to stabilize and promote the formation of thin gaseous disks. The role of modified gravitational potential/acceleration on disk formation in dwarf galaxies will be studied in detail in Hopkins et al. 2023 (in prep).

\begin{figure*}
    \centering
    \includegraphics[width=0.99\textwidth]{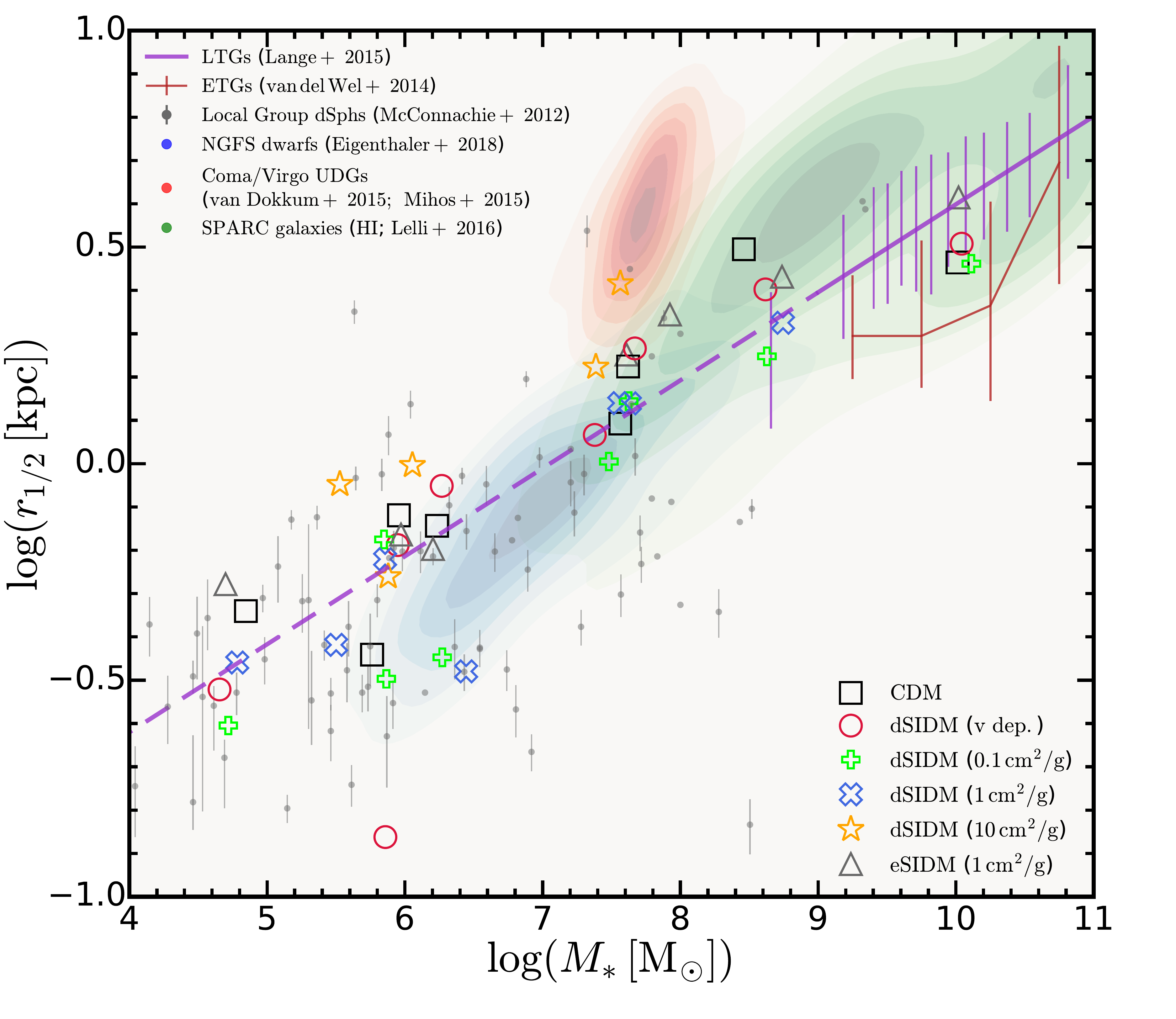}
    \caption{ \textbf{Size-mass relation of simulated (isolated) dwarf galaxies.} The stellar half-mass radius versus stellar masses of simulated dwarfs are shown with open markers (as labelled). We compare them with several observations of dwarf galaxies in the Local Universe: gray points with error bars, the Local Group dSphs compiled in \citet{McConnachie2012}; blue contours, the NGFS dwarfs in \citet{Eigenthaler2018}; green contours, the SPARC galaxies presented in \citet{Lelli2016,Lelli2016b}; red contours, the UDGs in the Coma and Virgo cluster from \citet{VanDokkum2015} and \citet{Mihos2015}; purple (red) line, the size-mass relation of the observed ``normal'' late-type (early-type) galaxies~\citep{Lange2015,vanderwel2014}. The simulated dwarfs are consistent with the median size-mass relation of LTGs in observations and its extrapolation. With mild dark matter self-interaction ( $(\sigma/m) \lesssim 1\cpm$), the sizes and masses of galaxies in general do not vary much from the CDM case. In some cases, the dSIDM models can produce compact dwarfs at $M_{\ast} \sim 10^{6}\msun$, in better agreement with Local Group observations. However, in the dSIDM-c10 model, dwarf galaxies have apparently more extended stellar content and are located at the diffuse end of the observed distribution.}
    \label{fig:mass-size}
\end{figure*}

\subsection{Stellar density profiles and star formation history}
\label{sec:baryonic-content-sdpro}

In Figure~\ref{fig:stellar-profile-classical} and Figure~\ref{fig:stellar-profile-bright}, we show the stellar density profiles of simulated classical and bright dwarfs, respectively. Each stellar density profile plot is paired with the plot of the archaeological star formation history of the galaxy. The star formation history is computed as the age distribution of stellar particles selected at $r\leq 10\%\,R^{\rm cdm}_{\rm vir}$ at $z=0$. In both classical and bright dwarfs, dSIDM with moderate cross-sections give rise to central stellar density profiles that are cuspier than the NFW profile and the galaxy stellar-half-mass-radii decrease correspondingly. These phenomena are likely caused by the more concentrated dark matter content in these dSIDM models. Similarly, the stellar profiles in eSIDM are cored due to the gravitational impact of thermalized dark matter cores. In the dSIDM-c10 model, the stellar distribution becomes cored and more extended, which is also related the decreased normalization of dark matter density profiles in this model. Specifically, in \citetalias{Shen2021}, coherent rotation of dark matter was found in the highly-dissipative models (including dSIDM-c10) and we showed that this drives halo deformation to oblate shapes. This combination of rotational support and change in shape actually leads to a {\em decline} in the central spherically-measured dark matter density, which we see here is reflected in the stellar distribution. In general, the compactness of the dark matter distribution appears to strongly influence the stellar density profile of dwarf galaxies. The star formation efficiency is regulated by the competition between feedback-driven ejection versus the gravitational attraction from dark matter~\citep[e.g.,][]{Grudic2020}. In equilibrium states (or after numerous cycles of star formation events), star formation is promoted (inhibited) in compact (diffuse) dark matter haloes. Stars formed before their dark matter halo is structurally modified (e.g.\ before a dark disky structure or a strong cusp owing to dissipation can form) can still relax with respect to the modified halo potential within a few dynamical time scales. One galaxy in this suite that deviates from the picture above is the classical dwarf m10v. In all dark matter models for this particular galaxy, cored stellar density profile are developed at the galaxy center, while the central stellar density increasing monotonously with dSIDM cross-section. The unique stellar content of m10v could be as a result of its distinct star formation history (bottom right panel of Figure~\ref{fig:stellar-profile-classical}), which is dominated by several recent (very low-redshift) starburst events. The system has therefore not yet relaxed from the perturbations of the recent star formation and feedback. 

In terms of the star formation history, m10q is clearly an early-forming dwarf with most of the star formation taking place at $z \gtrsim 2$ and a tail extended to $z\sim 0.7$. In dSIDM models with increasing effective cross-section, this tail of star formation ceases earlier, which is likely due to the faster depletion of star forming gas in the more compact dSIDM haloes. In m10v, despite drastically different star formation history from m10q, the recent peak of star formation also takes place earlier in dSIDM runs. Similarly, in the bright dwarf m11b, the star formation histories show apparent dips at low redshifts in dSIDM models, which do not occur in CDM and eSIDM runs. However, in m11a, the star formation histories in different dark matter models do not exhibit significant differences. 

\subsection{Galaxy size-mass relation}
\label{sec:baryonic-content-sizemass}

In Figure~\ref{fig:mass-size}, we compare the stellar-half-mass radii (as a function of stellar mass) of simulated dwarf galaxies (isolated ones only, do not include satellites of Milky Way-mass hosts) with observations of dwarf galaxies in the local Universe. These observations include the Local Group dwarf spheroidal galaxies (dSphs) compiled in \citet{McConnachie2012}, dwarf galaxies from the Next Generation Fornax Survey~\citep[NGFS,][]{Eigenthaler2018}, galaxies measured in the {\it Spitzer} Photometry and Accurate Rotation Curves~\citep[SPARC,][]{Lelli2016,Lelli2016b} project, and the ultra-diffuse galaxies (UDGs) in the Coma and Virgo cluster from \citet{VanDokkum2015} and \citet{Mihos2015}. The quoted effective radius (half-light radius) in literature has been converted to the half-mass radius assuming $r_{1/2}\simeq 4/3\, R_{\rm eff}$~\citep{Wolf2010}. The purple solid line shows the galaxy size-mass relation of ``normal'' late-type galaxies~\citep[LTGs;][]{Lange2015} and its extrapolation (purple dashed line) while the red solid line shows that of early-type galaxies~\citep[ETGs;][]{vanderwel2014}.

In general, despite some random galaxy-to-galaxy variations, the simulated dwarfs agree well with the observed dwarf population in the Local Universe, and follow the extrapolated size-mass relation of LTGs. The diversity of dwarfs is manifest as the distinction between LTGs and ETGs in massive sub-Milky Way-mass galaxies, the existence of UDGS and the large scatter in galaxy size at $M_{\ast} \sim 10^{8}\msun$ as well as the population of compact Local Group dwarfs that fall significantly lower than the median relation. With mild dark matter self-interaction ( $(\sigma/m) \lesssim 1\cpm$), galaxy sizes and masses in dSIDM or eSIDM do not vary much from the CDM case. This is consistent with previous FIRE-2 studies of dwarf galaxies in eSIDM~\citep{Robles2017,Fitts2019}. Some compact dwarfs at $M_{\ast} \lesssim 10^{7}\msun$ are found in dSIDM models, which are in better agreement with the observed compact dwarfs in the Local Group. However, the compact dwarf elliptical galaxies with large stellar masses ($M_{\ast} \gtrsim 10^{7}\msun$) in the Local Group~\citep[e.g.,][]{Tollerud2014,SGK2019} are still hard to produce in these isolated dwarf simulations, no matter which dark matter model is employed. This point will be revisited when we study the satellite galaxies of simulated Milky Way-mass hosts. Notably in the dSIDM-c10 model, simulated dwarfs exhibit systematically more extended stellar content and shift from the median relation. In this model, the bright dwarfs become more like analogs to UDGs and the classical dwarfs are located at the diffuse end of the observed distribution. The dSIDM-c10 model is therefore perhaps disfavored due to this systematic shift. However we caution that, as many of the observational studies above have noted, there could well exist a substantial population of even-lower-surface-brightness galaxies in nature which would simply not be detected given the present state-of-the-art surface brightness limits \citep[see][]{Wheeler2019}. The number of dwarfs in the simulation suite is too limited to tell if dSIDM with lower cross-sections are ruled out or are more consistent with the observed sample (in terms of the diversity of the stellar content).  

\begin{figure*}
    \centering
    \includegraphics[width=0.49\textwidth]{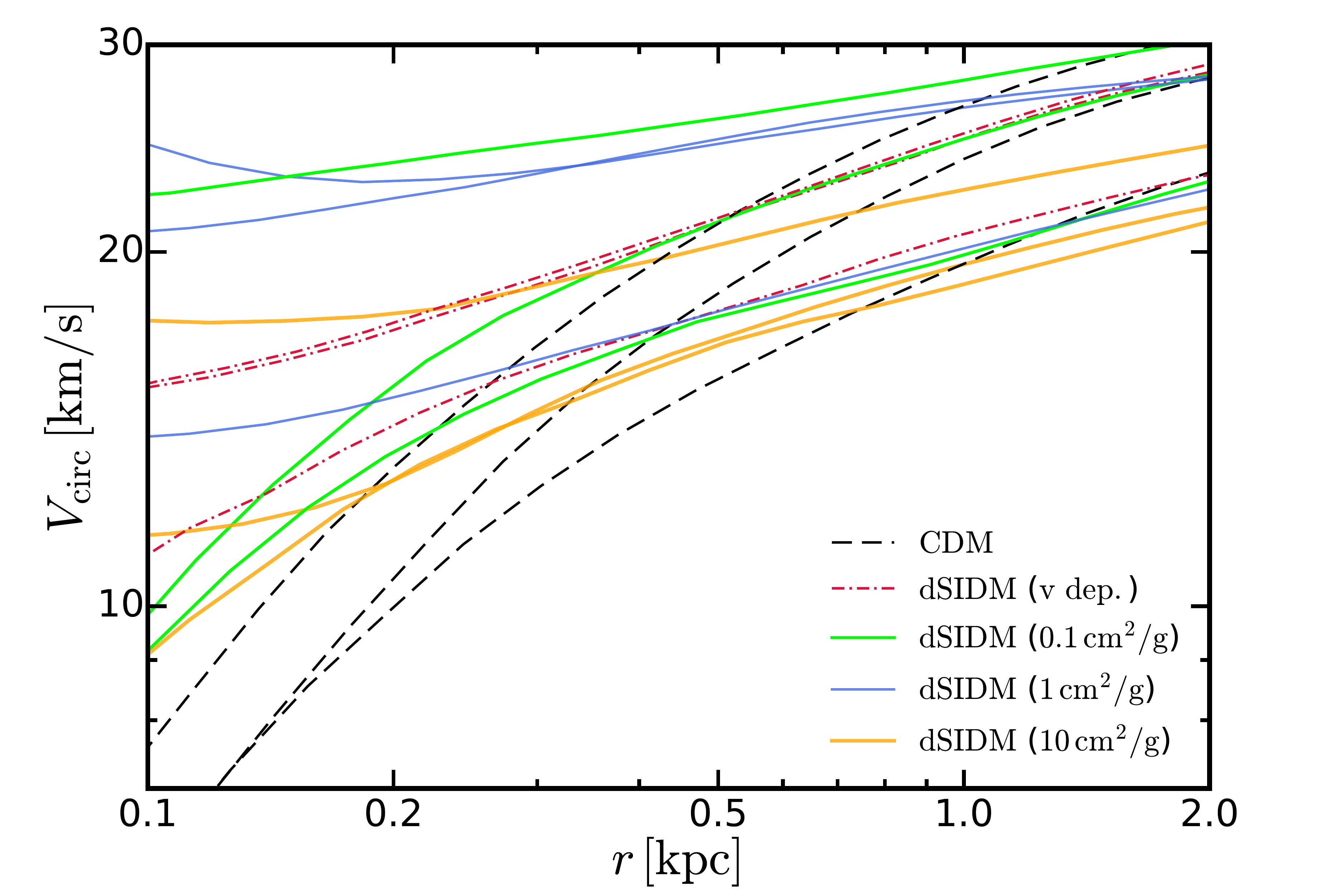}
    \includegraphics[width=0.49\textwidth]{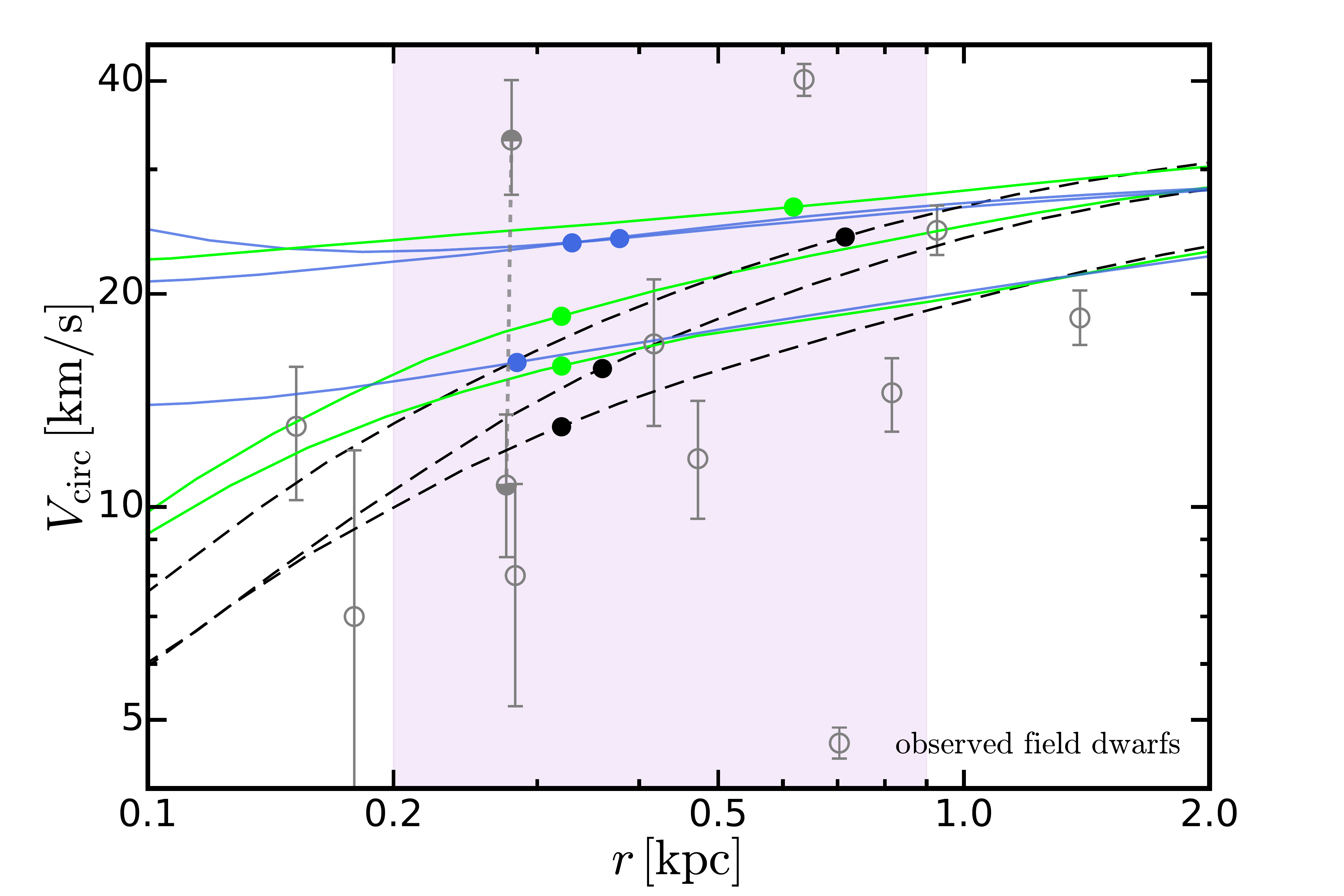}
    \includegraphics[width=0.49\textwidth]{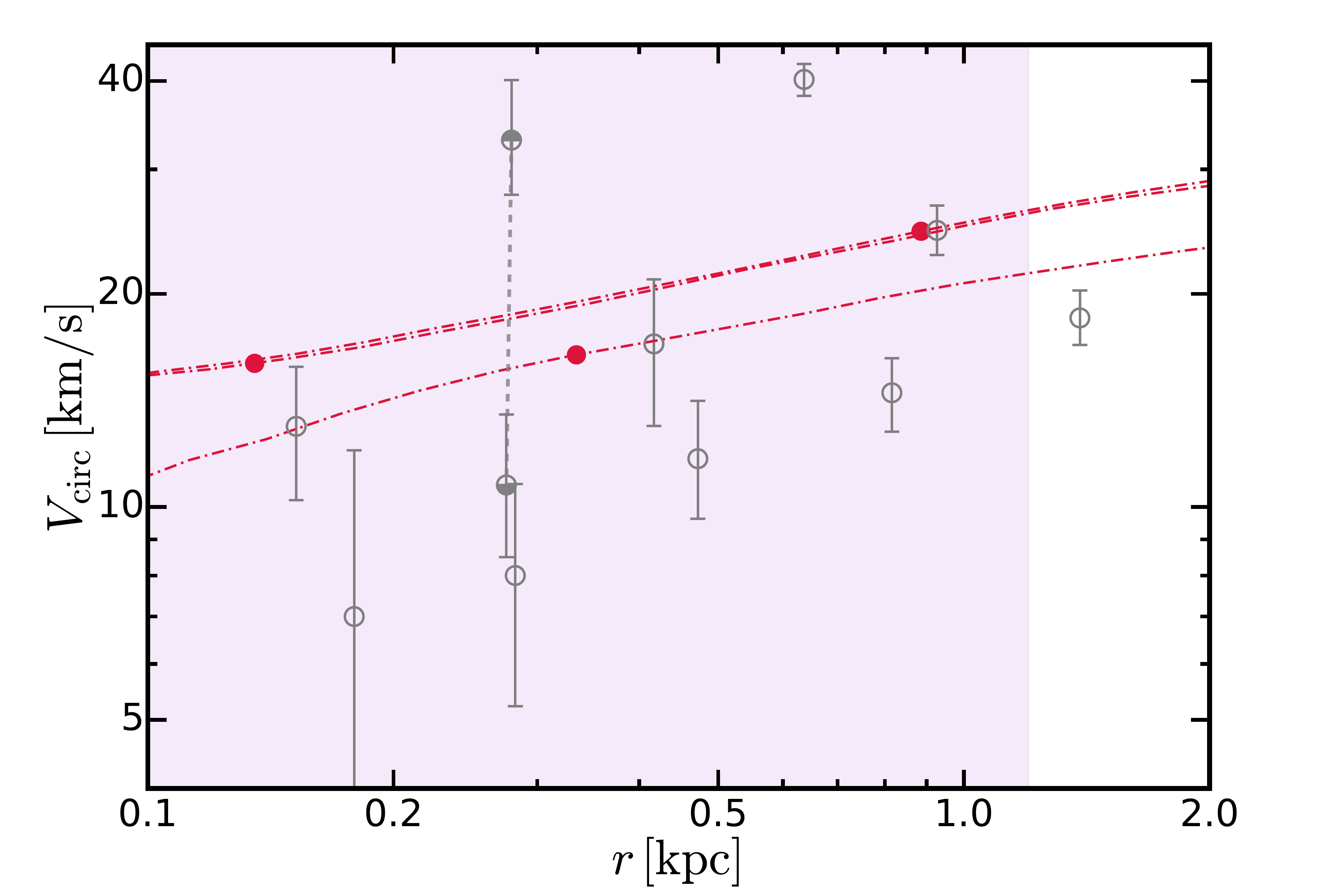}
    \includegraphics[width=0.49\textwidth]{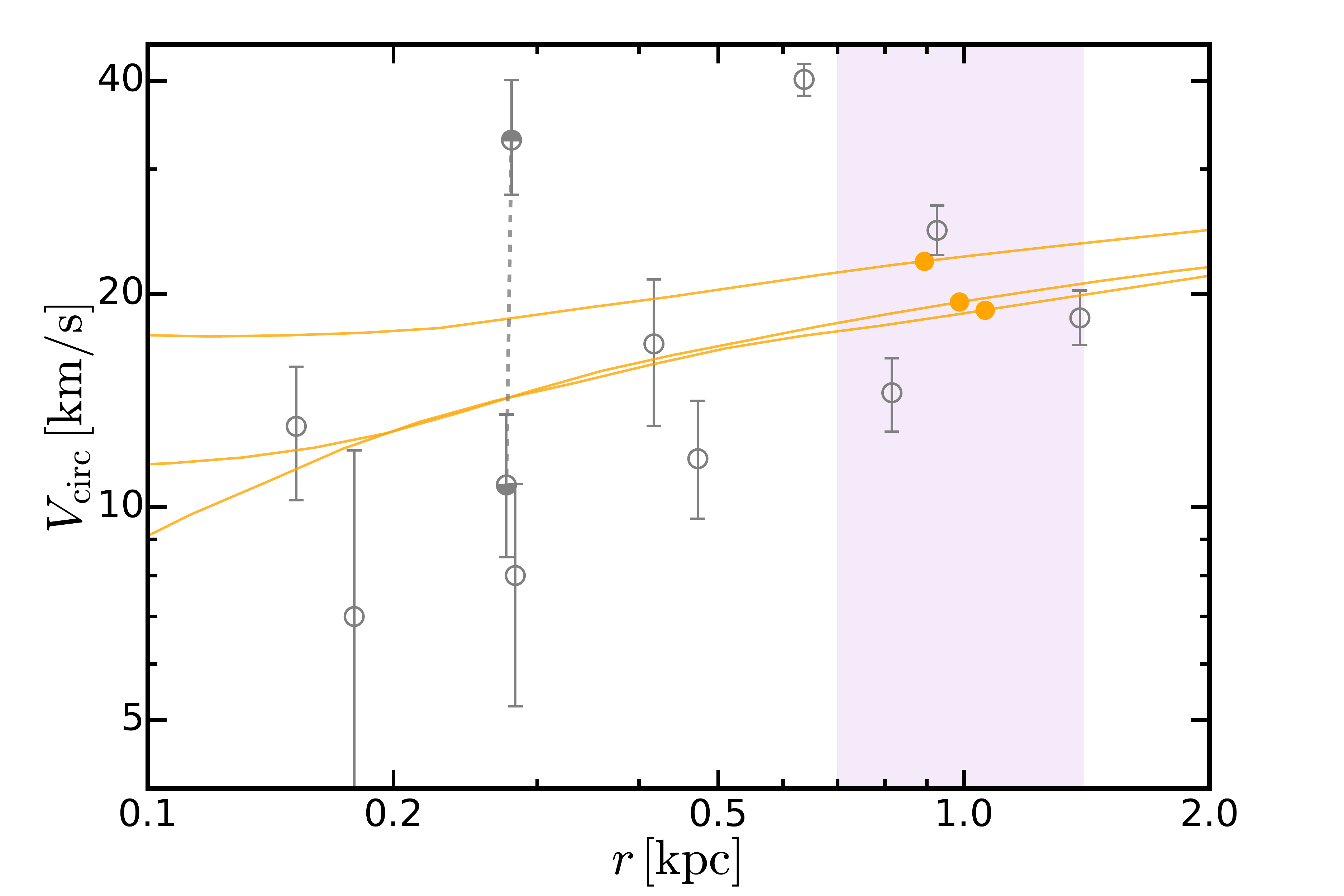}
    \caption{\textbf{Circular velocity profiles of simulated classical dwarfs compared with the observed field dwarfs in the Local Group.} {\it Top left:} Circular velocity profiles of the simulated dwarfs in different dark matter models. The circular velocities are enhanced at sub-kpc scale in dSIDM models. In the model with $(\sigma/m)=10\cpm$, the normalization of circular velocity profile decreases. {\it Top right:} We compare the results in CDM and dSIDM models with $(\sigma/m)=0.1/1\cpm$ with the observed field dwarfs in the Local Group (we show two measurements for Tucana, connected by a gray line; see text for details). The $r_{\rm 1/2}$ of these galaxies are shown by solid circles. We highlight the observed dwarfs of similar sizes to the simulated one ($0.2\kpc \lesssim r_{1/2} \lesssim 0.9\kpc$) with the purple shaded region. The CDM results are consistent with the majority of the observed dwarfs, but lower compared to the most compact dwarfs (NGC6822 and the older measurement of Tucana). The $\big(V_{\rm circ}(r_{\rm 1/2}), r_{\rm 1/2}\big)$ of these two dSIDM models are still marginally consistent with the observed dwarfs of similar sizes and improve the agreement for compact dwarfs. The circular velocities in the dSIDM models are about two times higher than the observed ones at small radii $r\lesssim 0.2\kpc$. {\it Bottom left:} We show the results of the velocity-dependent dSIDM model and compare them to the observed dwarfs with $0.1\kpc \lesssim r_{1/2} \lesssim 1.2\kpc$. {\it Bottom right:} We show the results of the model with $(\sigma/m)=10\cpm$ and compare them to the observed dwarfs with $0.7\kpc \lesssim r_{1/2} \lesssim 1.5\kpc$. The results from these two models are also consistent with observations.}
    \label{fig:rtcurve_field}
\end{figure*}

\begin{figure*}
    \centering
    \includegraphics[width=0.49\textwidth]{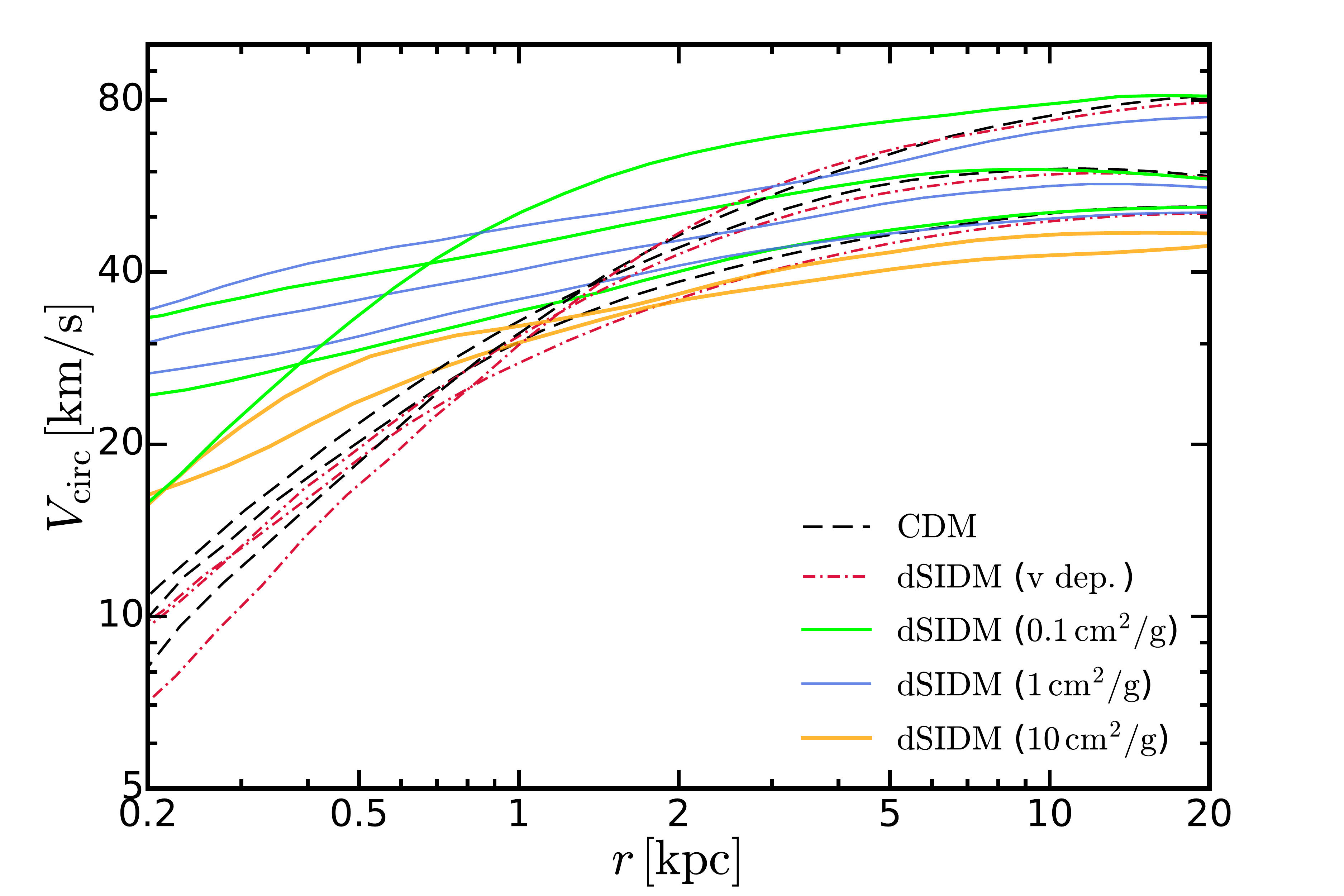}
    \includegraphics[width=0.49\textwidth]{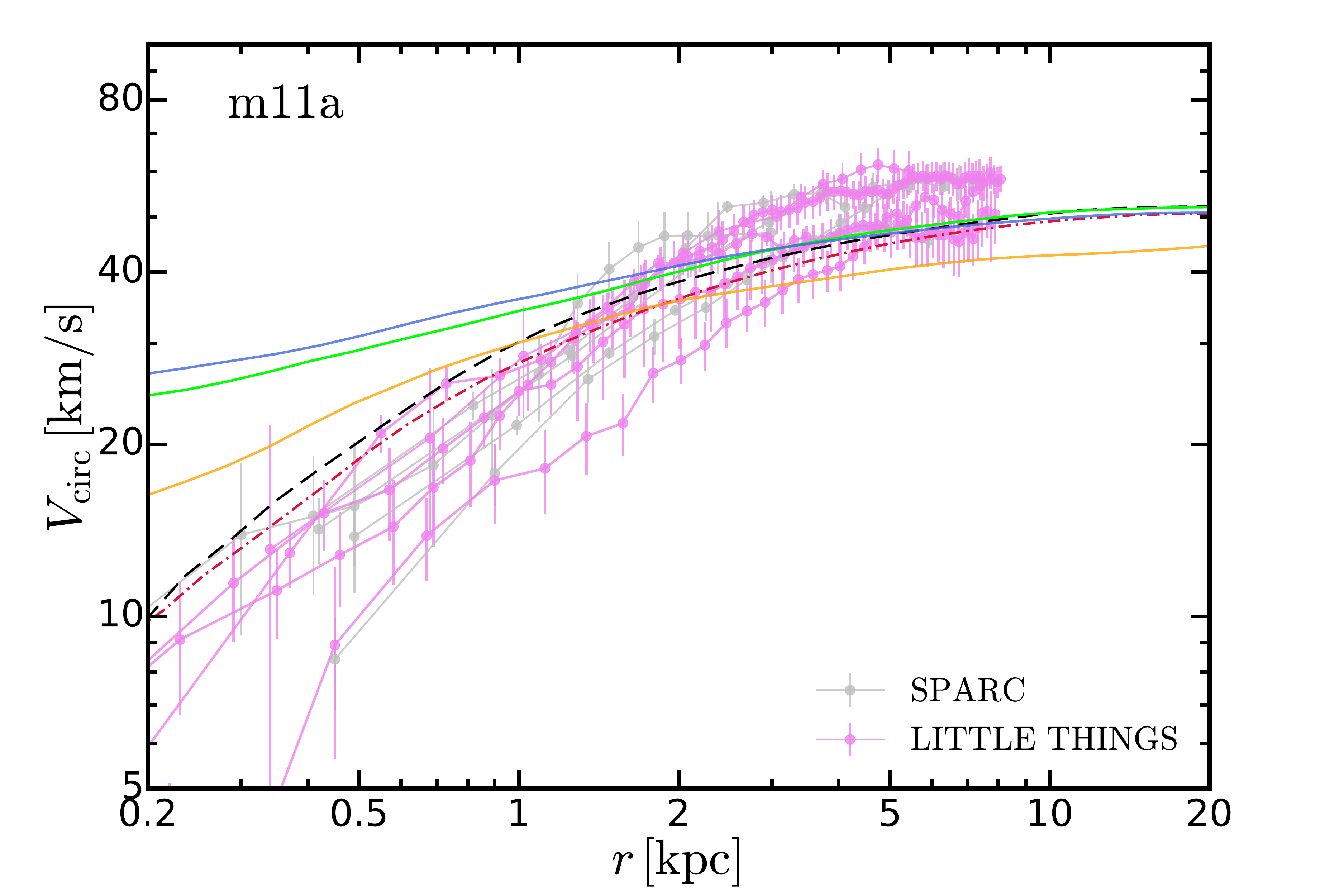}
    \includegraphics[width=0.49\textwidth]{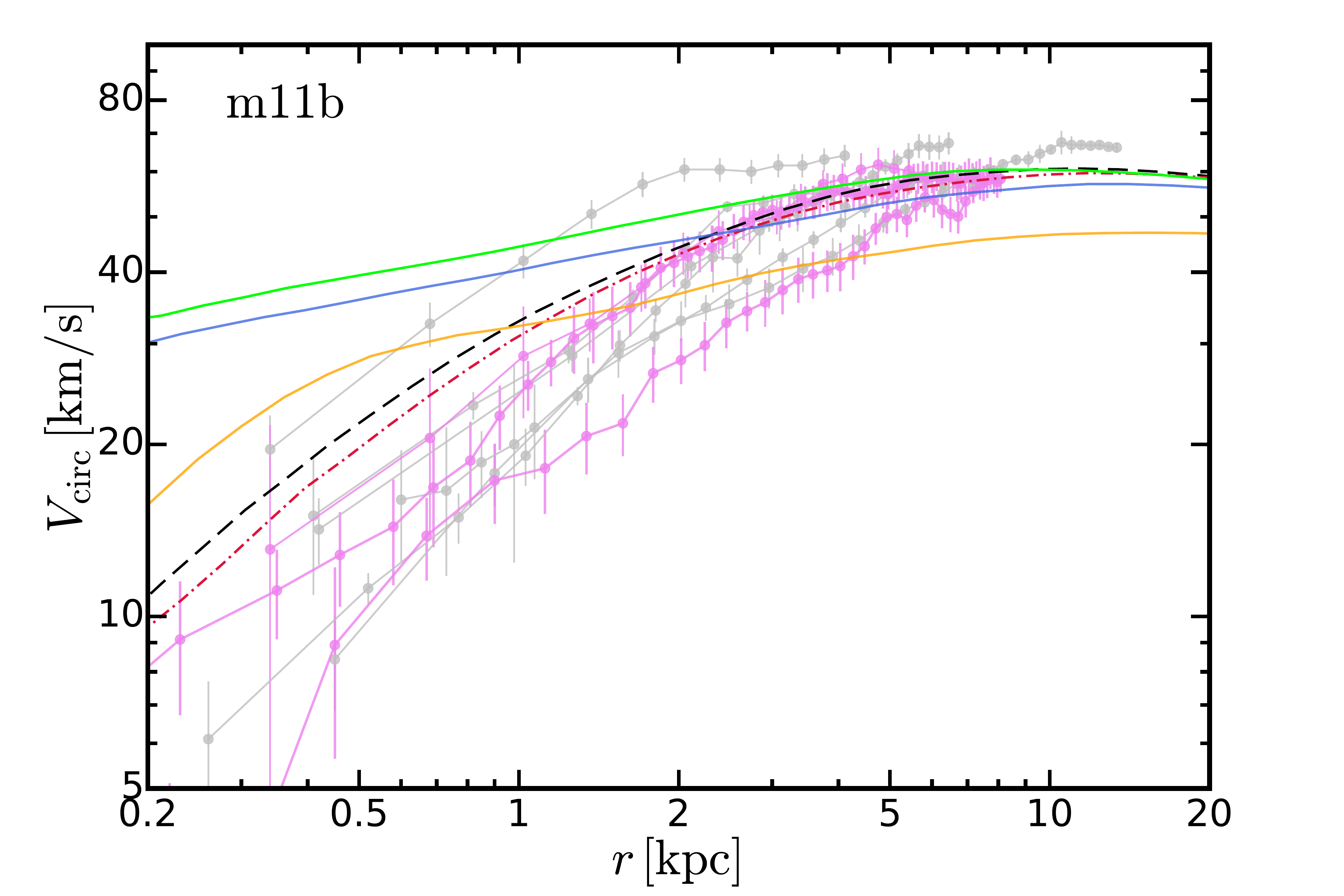}
    \includegraphics[width=0.49\textwidth]{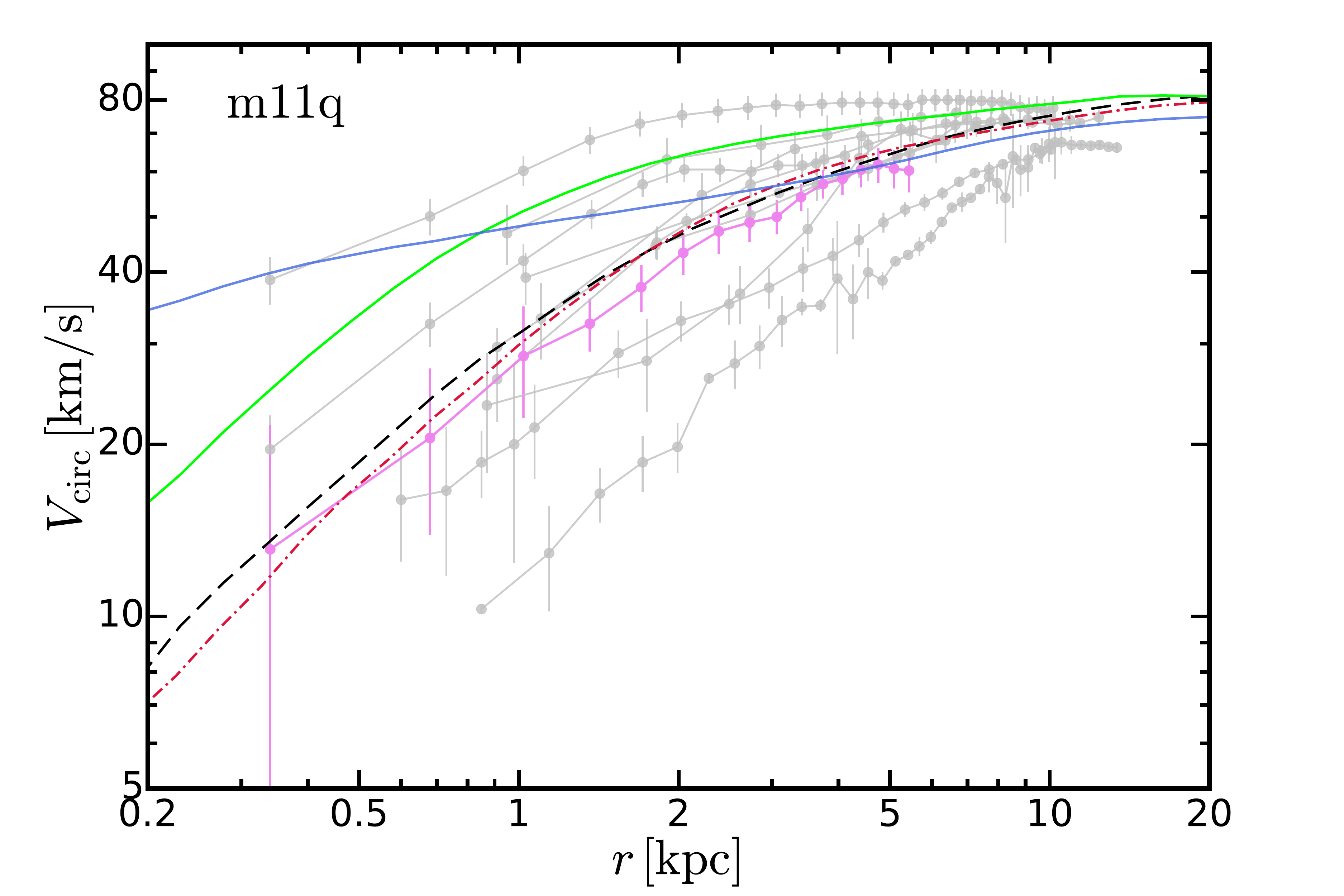}
    \caption{\textbf{Circular velocity profiles of simulated bright dwarfs compared with observed LSBs in the Local Universe.} {\it Right:} Circular velocity profiles of the bright dwarf galaxies in simulations. We compare the results with the measured circular velocities of LSBs observed in the field (see Section~\ref{sec:rotcurve} for details of the observed sample and selection criteria). Models with constant $\sigma/m$ that are consistent with in the classical dwarfs (with low $V_{\rm c}$) generally produce too concentrated galaxies at high $V_{\rm c}$, but the velocity-dependent model is consistent over the entire range we consider here.}
    \label{fig:rtcurve_lsbs}
\end{figure*}

\section{Galaxy circular velocity profiles}
\label{sec:rotcurve}

In this section, we will compare the circular velocity profiles of the simulated dwarfs with observations and attempt to derive constraints for dSIDM. First, we will analyze the isolated dwarfs (main ``target'' haloes in simulations). The ideal observational counterparts for the simulated classical dwarfs (m10 galaxies, see Table~\ref{tab:sim}) are the observed field dwarfs in the Local Group (with distances to the Milky Way and M31 $d>300\kpc$). These field dwarfs typically have sub-kpc $r_{\rm 1/2}$ and $M_{\ast} \lesssim 10^{7}\msun$, which are comparable to the m10 galaxies. The observational counterparts for the simulated bright dwarfs (m11 galaxies) are the LSBs in the Local Universe, usually with $r_{\rm 1/2}$ of several kpc and $10^{7}\msun \lesssim M_{\ast} \lesssim 10^{9}\msun$. In addition to the isolated dwarfs, we will analyze the subhaloes (and the satellite galaxies they host) of the simulated Milky Way-mass hosts (m12 galaxies) and compare them to the observed satellites of the Milky Way and M31.

\subsection{Observational samples}
\label{sec:obdata}

For satellite galaxies, we adopt the Milky Way and M31 satellites compiled in \citet{SGK2019}, which was updated based on the \citet{McConnachie2012} compilation. These dwarf galaxies were classified as satellites with their distances to the Milky Way or M31 smaller than $300\kpc$ (following the criterion adopted in \citet{Wetzel2016} and \citet{SGK2019}). For the Milky Way satellites, the dSphs presented in \citet{Wolf2010} were included and the implied circular velocity at the three-dimensional (de-projected) half-mass radius, $V_{\rm 1/2} = V_{\rm circ}(r_{\rm 1/2})$, has been calculated using their formula based on the average velocity dispersion of stars. In addition, the H\Rmnum{1}-based circular velocity measurement of the Small Magellanic Cloud (SMC) from \citet{Stanimirovic2004} and the proper motion-based circular velocity measurement of the Large Magellanic Cloud (LMC) from \citet{vanderMarel2014} were included. For the satellites of M31, the compilation included the $r_{\rm 1/2}$ and $V_{\rm 1/2}$ measurements from \citet{Tollerud2014}. For the dwarfs in the Local Field (with distances to hosts larger than $300\kpc$), the compilation included the $r_{\rm 1/2}$, $V_{\rm 1/2}$ and $\sigma_{v,\ast}$ from \citet{Kirby2014} where possible, though with modifications to the three galaxies with evidence of rotation~\citep{SGK2014}. A recent measurement~\citep{Taibi2020} on the field dwarf ``Tucana'' obtained a much lower dynamical mass of the system than the previous measurements~\citep{Fraternali2009,Gregory2019}, so we update the compilation correspondingly.

For the LSBs, we adopt the H\Rmnum{1} rotation curves and mass models from the ``Local Irregulars That Trace Luminosity Extremes, The H\Rmnum{1} Nearby Galaxy Survey''~\citep[LITTLE THINGS,][]{Oh2015}. The mass modelling results in \citet{Oh2015} showed that the selected galaxies have a typical halo mass of $\sim 10^{10-11} \msun$ and stellar mass of $\sim 10^{7-9} \msun$, which are in good agreement with the simulated bright dwarfs. In addition, we include the H\Rmnum{1}/H$\alpha$ rotation curves and mass models from the ``{\it Spitzer} Photometry and Accurate Rotation Curves''~\citep[SPARC,][]{Lelli2016} project. Given the limited statistics provided by only three simulated bright dwarfs, we will do a case-by-case comparison by selecting observed galaxies based on their maximum circular velocities, effective radii and inferred stellar masses.

\subsection{Circular velocity profiles of isolated dwarfs}
\label{sec:rotcurve_iso}

In the top left panel of Figure~\ref{fig:rtcurve_field}, we show the circular velocity profiles of the simulated classical dwarfs in different dark matter models. In general, the circular velocities at $r\lesssim 1\kpc$ increase in dSIDM models with $0.1\cpm \lesssim (\sigma/m)_{\rm eff} \lesssim 1\cpm$~\footnote{At the mass scale of classical dwarfs, the effective cross-section $(\sigma/m)_{\rm eff}$ is about $0.3\cpm$, where $(\sigma/m)_{\rm eff}$ follows the definition in \citetalias{Shen2021}.} and the circular velocity profiles are almost flat at the center. For example, the circular velocities at $r\simeq 0.2\kpc$ are enhanced by about a factor of two in the dSIDM-c1 model compared to the CDM case. This is a direct consequence of the cuspy central density profiles in dSIDM models, as detailed in \citetalias{Shen2021}. In the dSIDM-c10 model, circular velocity profiles have similar flat shapes but with systematically lower normalizations than the dSIDM-c1 model, which is likely related to the coherent rotation and halo deformation in the strong dissipation limit. 

In the other three panels of Figure~\ref{fig:rtcurve_field}, we show specifically the $(V_{\rm 1/2},\,r_{\rm 1/2})$ of the simulated dwarfs and compare them to the circular velocities of $10$ observed Local Group field dwarfs (compiled in Section~\ref{sec:obdata}). For Tucana, both the recent measurement~\citep{Taibi2020}, which attempts to subtract a potential correction (still somewhat uncertain) for unresolved stellar binaries, and an older measurement \citep{Fraternali2009} without such a correction are shown and linked by a gray dashed line in the figure. The circular velocity profiles in CDM are consistent with the bulk of the observed dwarfs, except two dense outliers (Tucana, if we take the older measurement, and NGC6822) with $V_{\rm 1/2}\sim 30\operatorname{-}40 \kms$. The dSIDM-c0.1 and dSIDM-c1 models are marginally consistent with observations: the $V_{\rm 1/2}$ of some simulated dwarfs are slightly higher than the observed dwarfs of similar $r_{\rm 1/2}$ except for NGC6822 (if we adopt the new measurement of Tucana) but the differences at this level are not enough to rule out these models given the limited statistics. For the velocity-dependent dSIDM model, the simulated dwarfs are consistent with the relatively compact observed dwarfs but may be in tension with the six diffuse ones. Again the limited statistics prevent us from drawing any conclusions about the model. For the dSIDM-c10 model, although the circular velocities at small radii appear higher the observed ones, the $V_{\rm 1/2}$ are still consistent with the observed dwarfs with comparable sizes. The potential problem with this model is that the stellar content of all simulated dwarfs is relatively diffuse, and the range of galaxy stellar effective radii may not be diverse enough to match observations. We also note that there is one observed galaxy (NGC6822, or two if the older Tucana measurement is used) lying above the circular velocity profiles of any simulated galaxies regardless of the dark matter model employed. Even the model with the highest degree of dissipation used here cannot produce analogs of these compact systems. If the discrepancy is real (not the result of e.g. unresolved binaries or other sources of dispersion), the physical origin of these systems in the field is still a challenge to existing cosmological simulations~\citep[e.g.,][]{Dutton2016, Fattahi2016, Sawala2016, Wetzel2016, SGK2019}.

In the top left panel of Figure~\ref{fig:rtcurve_lsbs}, we show the circular velocity profiles of the simulated bright dwarf galaxies in different dark matter models. Circular velocities in the dSIDM-c0.1 and dSIDM-c1 models are enhanced to about $30-40\kms$ at $r \lesssim 1\kpc$. The circular velocity profiles in the dSIDM-c10 model have similar shapes but lower normalizations. Those in the velocity-dependent dSIDM model are almost indistinguishable from the CDM case, due to the limited effective cross-sections in the bright dwarfs. In the other three panels, we compare the circular velocity profiles of each simulated dwarf with the H\Rmnum{1}-based measurements from the LITTLE THINGS survey \citep{Oh2015} and the SPARC survey \citep{Lelli2016} as introduced in Section~\ref{sec:obdata}. For m11a, we select observed galaxies with maximum circular velocities $40\kms \lesssim V^{\rm max}_{\rm circ} \lesssim 60\kms$ and, for m11b, we select observed galaxies with $50\kms \lesssim V^{\rm max}_{\rm circ} \lesssim 70\kms$. In addition, for both galaxies, we require the observed sample to have $0.5\kpc \lesssim r_{\rm 1/2} \lesssim 3\kpc$ and $10^{7}\msun \lesssim M_{\ast} \lesssim 10^{8.5}\msun$. From these comparisons, we find that the CDM and the velocity-dependent dSIDM model are fully consistent with observations at these mass scales. However, the circular velocities in the dSIDM-c0.1 and dSIDM-c1 models are about two times higher than the observed values at sub-kpc scale, and the discrepancy appears to be larger than both the observational uncertainties as well as the galaxy-to-galaxy scatter. 

\begin{figure*}
    \centering
    \includegraphics[width=1.02\textwidth]{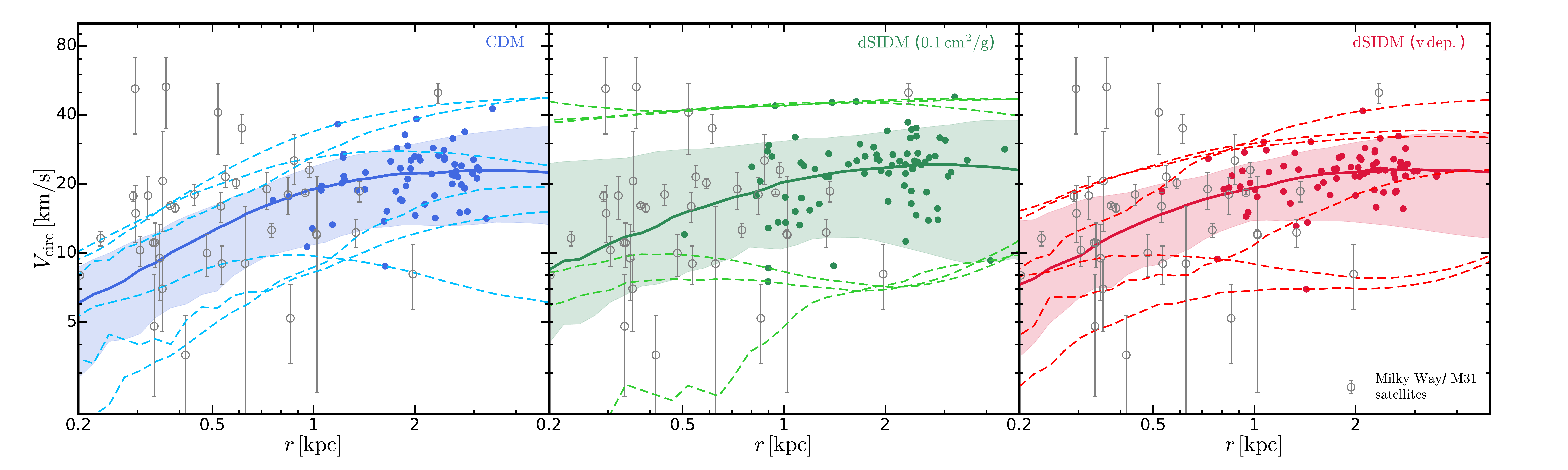}
    \caption{\textbf{Circular velocity profiles of satellite galaxies of simulated Milky Way-mass galaxies and compared with observations.} The circular velocity profiles in each dark matter model are shown in each column respectively. The solid lines and the shaded regions show the median and the $1.5\sigma$ scatter ($86\%$ of the sample) of the curves. The dashed lines highlight the three circular velocity profiles with the highest (and the three with the lowest) circular velocities at $r=1\kpc$. Gray circles with error bars show the $(V_{\rm 1/2},\, r_{\rm 1/2})$ of observed Milky Way and M31 satellites compiled in Section~\ref{sec:obdata}. The $r_{\rm 1/2}$'s of simulated satellites are marked by solid circles. The identified subhaloes in  simulations are selected as satellites if they have galactocentric distance $20\kpc < d < 300\kpc$, and with at least $200$ dark matter particles and $10$ associated stellar particles (equivalently $M_{\ast} \geq 10\,m_{\rm b}$). The selected satellites are in the mass range $M_{\ast} \sim 10^{5} \operatorname{-} 10^{8}\msun$, in concordance with the observed sample. The circular velocity profiles in different models are almost indistinguishable compared to the scatter among the observed satellites, despite the slightly larger median rotation velocities and upper scatter in the dSIDM-c0.1 model. Circular velocity profiles from all three models are consistent with the bulk of the observed dwarfs, although the predicted galaxy sizes are systematically larger. The smallest $r_{\rm 1/2}$ reached in the two dSIDM models is smaller than the CDM case, down to about $\sim 500\,{\rm pc}$. As indicated by the dashed lines, the most compact satellites in the dSIDM-c0.1 model agree better with the observed compact dwarfs in the Local Group, though the stellar content is still puffier compared to observations.}
    \label{fig:rtcurve_sate}
\end{figure*}

For the massive dwarf m11q, we select observed galaxies with $60\kms \lesssim V^{\rm max}_{\rm circ} \lesssim 80\kms$, $1\kpc \lesssim r_{\rm 1/2} \lesssim 5\kpc$ and $10^{8}\msun \lesssim M_{\ast} \lesssim 10^{9}\msun$. The CDM and the velocity-dependent dSIDM models are again consistent with the median circular velocity profiles of the observed dwarfs. However, due to the prominent diversity of the observed circular velocity profiles at the mass scale, the dSIDM-c0.1 and dSIDM-c1 models are still marginally consistent with observations. 

In conclusion, the comparisons of the three bright dwarfs with observations appears to disfavor both the constant cross-section dSIDM models with $(\sigma/m) \gtrsim 0.1\cpm$. However, a velocity-dependent model is still viable to produce unique phenomena in lower mass dwarfs while maintaining consistency with the H\Rmnum{1}-based observations of bright dwarfs. 

One important caveat we note is that the measurements here all adopt H\Rmnum{1} as the kinematic tracer of the gravitational potential. This certainly involves an additional layer of uncertainties in fitting the H\Rmnum{1} velocity field and asymmetric drift corrections. In addition, the galaxies selected in the observational sample all are chosen to exhibit cold dense gas disks. Most galaxies so selected are morphologically spiral or irregular galaxies, and the observed samples by construction will miss elliptical or spheroidal dwarf galaxies lacking a dense H\Rmnum{1} disk, which some authors have argued may be more compact than the late-type disky galaxies of similar stellar masses \citep[e.g.,][]{vanderwel2014,Eigenthaler2018}. This could potentially bias the comparison here and naively might loosen the constraints on dSIDM models. However, as shown in Figure~\ref{fig:gas_image}, disk-like structures of cold neutral gas are indeed prominent in m11a and m11b and are promoted in dSIDM models. Therefore, compared to the CDM case, galaxies in dSIDM models would be more likely to appear in H\Rmnum{1} selected samples in observations, but actually match less well with the measured circular velocity profiles of those samples.

\begin{figure*}
    \centering
    \includegraphics[width=0.98\textwidth]{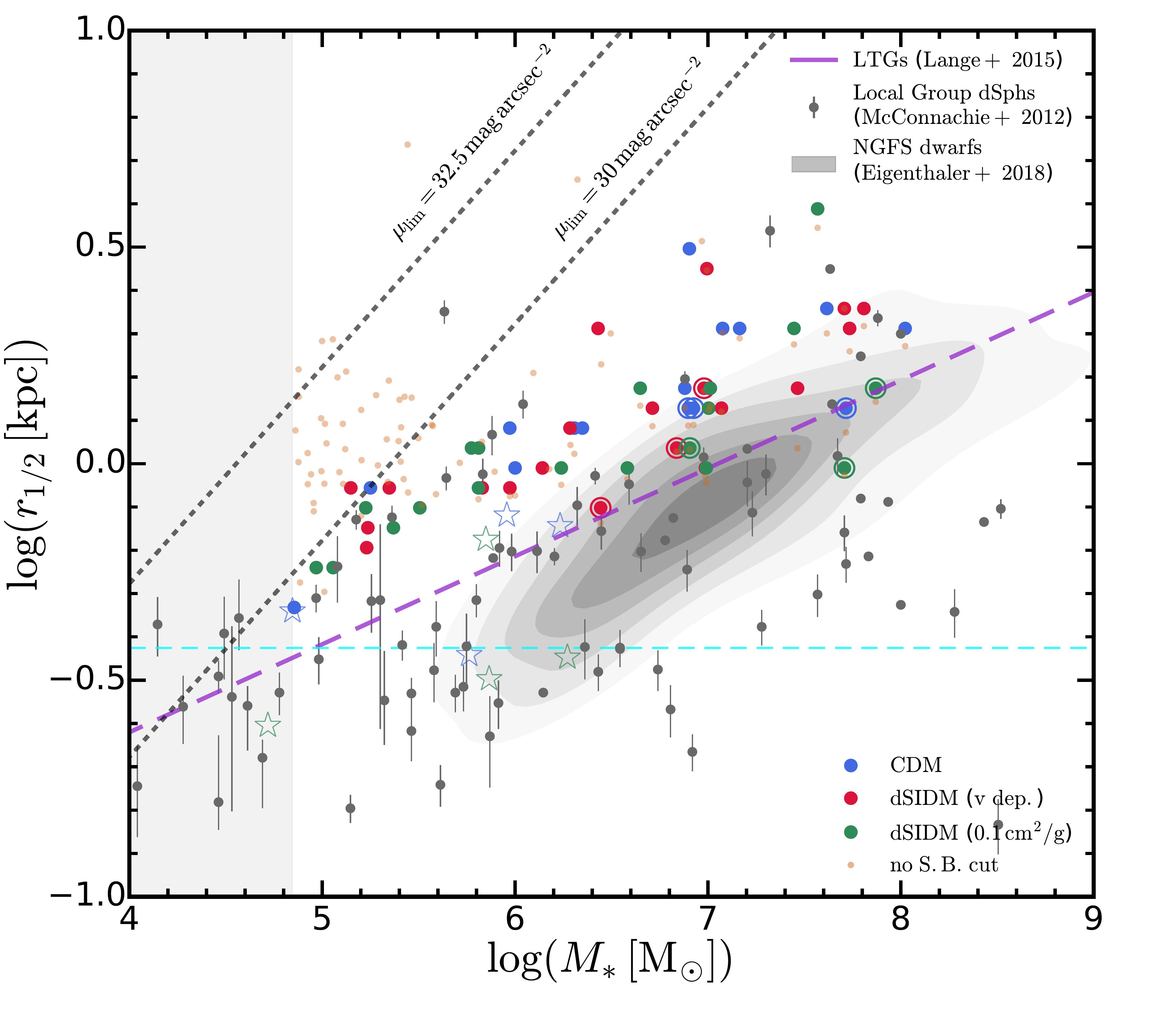}
    \caption{\textbf{Size-mass relation of satellite galaxies.} We show the stellar-half-mass-radius versus stellar mass of satellites of the simulated Milky Way-mass host(s). Only the high-resolution runs are considered here. The satellites from simulations follow the same selection criteria as in Figure~\ref{fig:rtcurve_sate}. The solid points show satellite sizes corrected for the surface brightness limit in observations. The black dotted lines indicate the surface brightness limit $30\,{\rm mag}\,{\rm arcsec}^{-2}$ for the SDSS surveys and the limit with an order of magnitude increasing sensitivity. For reference, the Local Group dwarfs~\citep{McConnachie2012} are shown by gray points and the NGFS dwarfs~\citep{Eigenthaler2018} are shown by the gray shaded contours. The purple dashed line is the extrapolation of the size-mass relation of local late-type galaxies~\citep{Lange2015}. The left shaded region indicates the mass resolution limit of the simulated satellites. The horizontal cyan dashed line indicates radius limit where the enclosed dark matter particle number is $\leq 200$ for a typical satellite central density $\rho_{\rm dm} \simeq 10^{7.5}\msun\,\kpc^{-3}$. The markers encircled highlight the three most compact dwarfs (with highest rotation velocities at $r=0.5\kpc$) in each run. A significant population of low-mass satellites in simulations are not detectable in current observations. For those in the observed regime, no obvious difference is found between CDM and dSIDM models. Massive satellites in dSIDM models are slightly more compact than their CDM counterparts, but they are still systematically puffier than the observed ones. In all the models, the satellites with the most compact dark matter content (highest circular velocities identfied in Figure~\ref{fig:rtcurve_sate}) also have the most compact stellar content. However, despite similar stellar masses, they have about three times larger $r_{1/2}$ than the observed compact dwarf elliptical galaxies. For reference, the $(r_{1/2},M_{\ast})$'s of simulated classical dwarfs (isolated systems) are shown as open stars. With an order-of-magnitude better mass resolution, the isolated dwarfs have slightly more compact stellar content that is in better agreement with the observed samples. This hints the resolution-dependent uncertainties, which will be discussed in Appendix~\ref{app:resolution}.}
    \label{fig:sat_rhalf}
\end{figure*}

\subsection{Circular velocity profiles of satellites of Milky Way-mass hosts}
\label{sec:rotcurve_sat}

The comparisons above focus on isolated systems to avoid contamination with environmental effects, but the derived constraints are subjected to galaxy-to-galaxy statistical variations given the limited number of isolated dwarfs in the simulation suite. An alternative way to constrain the dSIDM models is to compare satellite galaxies of more massive hosts to improve the statistics. For this purpose, we analyze the three low-resolution runs of Milky Way-mass hosts (m12i, m12m and m12f) and a high-resolution run for m12i (details listed in Table~\ref{tab:sim}). Their subhaloes (as well as the associated stellar content) are identified with the procedure introduced in Section~\ref{sec:sim_substructure}.

In Figure~\ref{fig:rtcurve_sate}, we show the circular velocity profiles of satellite galaxies of simulated Milky Way-mass galaxies (the median curve, the $1.5\sigma$ scatter and the three satellites with the maximum/minimum rotation velocities at $r=1\kpc$) and compare them with the observed satellites of the Milky Way and M31 compiled in \citet{SGK2019} as introduced in Section~\ref{sec:obdata}. For simulations, the identified subhaloes are classified as ``satellites'' if their distance from the center of the Milky Way or M31-analog is $20\kpc \leq d \leq 300\kpc$. We only keep satellites with dark matter particle number $N_{\rm dm}\geq 200$ and associated stellar particle number $N_{\ast}\geq 10$, which roughly corresponds to a stellar-mass cut of $M_{\ast}\geq 7\times 10^{4} \,(5.6\times 10^{5})\,\msun$ for high-(low-)resolution simulations. For reference, the minimum stellar mass of the observed sample we select is $7.3\times 10^{4}\msun$ ($3\times 10^{5}\msun$) for M31 (Milky Way). As shown in Figure~\ref{fig:rtcurve_sate}, circular velocities in dSIDM models are slightly higher than their CDM counterparts (both the median and upper scatter), but the differences are subdominant compared to intrinsic scatter of the observed satellites. Despite the fact that the circular velocity profiles in all the models are consistent with the majority of the observed dwarfs, the stellar-half-mass radii are systematically larger than the observed values. This will be discussed in more details in the comparison of the size-mass relation below. In addition, the CDM and the velocity-dependent dSIDM model fail to produce the most compact dwarf with $r_{\rm 1/2}\lesssim 1\kpc$ and $V_{\rm 1/2}\sim 40\kms$, which are typically elliptical or irregular galaxies in the M31 subgroup with stellar masses $\gtrsim 10^{8}\msun$. However, the dSIDM-c0.1 model gives larger scatter in the rotation velocites at sub-kpc scale and can produce analogs of those galaxies. But we need to note that the presence of compact satellite analogs in the dSIDM run (while not in CDM) needs further validation with improved statistics of the host systems simulated (at this point, it is difficult to say how significant the result is).

The typical mass and size of the satellites studied here are similar to the isolated classical dwarfs studied in Section~\ref{sec:rotcurve_iso}. However, the differences between dark matter models found in these satellites are smaller than what we found for field dwarfs. First, this could be related to additional factors that affect galaxy structure in a group environment, such as dynamical friction, tidal and ram pressure stripping~\citep[e.g.,][]{Quinn1986,Colpi1999,Taylor2001,Zentner2003,Gan2010,Jiang2016}. But a more plausible explanation would be resolution effects. The Milky Way-mass host simulations are about $30$ times poorer in mass resolution (i.e. $m_{\rm b}=7000\msun$ for m12i versus $250\msun$ for m10q) than the isolated dwarf simulations. Since the impact of dSIDM typically shows up at very small radii $r\lesssim 500\,{\rm pc}$, this could be challenging to resolve in $m_{\rm b} = 7000\msun$ runs (see the convergence plots of m10q and m10v in \citealt{Hopkins2018}). 

In Figure~\ref{fig:sat_rhalf}, we show the size-mass relation for selected satellite galaxies from simulations and compare them to observations. The Local Group dSphs and the NGFS dwarfs compiled for Figure~\ref{fig:mass-size} are shown here again for reference. The galaxy size measurements are often affected by the surface brightness detection limit in observations. Following \citet{Wheeler2019}, this is estimated to be $\mu_{\rm V} = 30 \mmag\,{\rm arcsec}^{-2}$ for SDSS, which corresponds to a physical stellar surface density limit $\Sigma^{\rm lim}_{\ast} = 0.036 \msun\,{\rm pc}^{-2}$ assuming solar absolute magnitude $M_{\odot,{\rm V}} = 4.83$ and a stellar mass-to-light ratio of $M_{\ast}/L \simeq 1 M_{\odot}/L_{\odot}$. The limit is indicated with the black dotted line in Figure~\ref{fig:sat_rhalf} when $\Sigma_{1/2} \equiv M_{\ast}/\pi\,r^{2}_{1/2} = \Sigma^{\rm lim}_{\ast}$. We also show the surface density limit with an order of magnitude increasing sensitivity at $\mu_{\rm V} = 32.5 \mmag\,{\rm arcsec}^{-2}$ for future surveys. In simulations, a significant population of low-mass satellites have surface brightness close or below the observational detection limit, the majority of which will not be detected in current surveys. Even the bright ones are potentially affected by the surface brightness cut in size/mass measurements. To correct for this effect, we measure the stellar surface density profile and truncate it where the average enclosed stellar surface density drop below $\Sigma^{\rm lim}_{\ast}$. The stellar mass is then corrected to the enclosed stellar mass within the cut-off radius and the $r_{1/2}$ is also corrected correspondingly. If the stellar surface density of the satellite is too low to identify the cut-off radius, the satellite is removed from the sample. After this correction, most of the satellites eventually reside in the detectable region on the size-mass plane. However, compared to the observed satellites, they are systematically more diffuse which is consistent with what we found in Figure~\ref{fig:rtcurve_sate}. No obvious difference between dark matter models is found, despite the fact that satellites at the massive end in the dSIDM-c0.1 model are more compact than the CDM counterparts. It is usually the satellites with the most compact dark matter content (highest circular velocities at sub-kpc scale) that also exhibit the most compact stellar content. In Figure~\ref{fig:rtcurve_sate}, we found that the most compact satellites in the dSIDM-c0.1 model are better counterparts to the observed compact dwarf elliptical galaxies in the Local Group, in terms of their circular velocities. However, in the size-mass plane, it is clear that these satellites found in simulations still do not have compact enough stellar content to match the most compact observed systems. This discrepancy could owe to observational effects (e.g.\ selection effects making it much easier to identify high-surface-brightness objects, or the fact that observations often use the light-weighted, Sersic-estimated profiles rather than the mass-weighted $r_{1/2}$ we measure here), or to the fact that some ``satellites'' may have their light profiles dominated by a single, massive/compact star cluster (or even themselves be a star cluster entirely rather than a true dwarf galaxy) as shown in some very high-resolution simulations in \citet{ma:fire.reionization.epoch.galaxies.clumpiness.luminosities}. Exploring these possibilities will require more detailed forward-modeling in future work. Similar to the point we made above about the circular velocity profiles of satellites, we caution that this discrepancy could also be a resolution effect. Specifically, with about an order-of-magnitude better mass resolution, the simulated isolate dwarfs at a similar mass scale are in better agreement with the observed samples. The potential resolution effects will be discussed in detail in Appendix~\ref{app:resolution}.

\begin{figure*}
    \centering
    \includegraphics[width=0.49\textwidth]{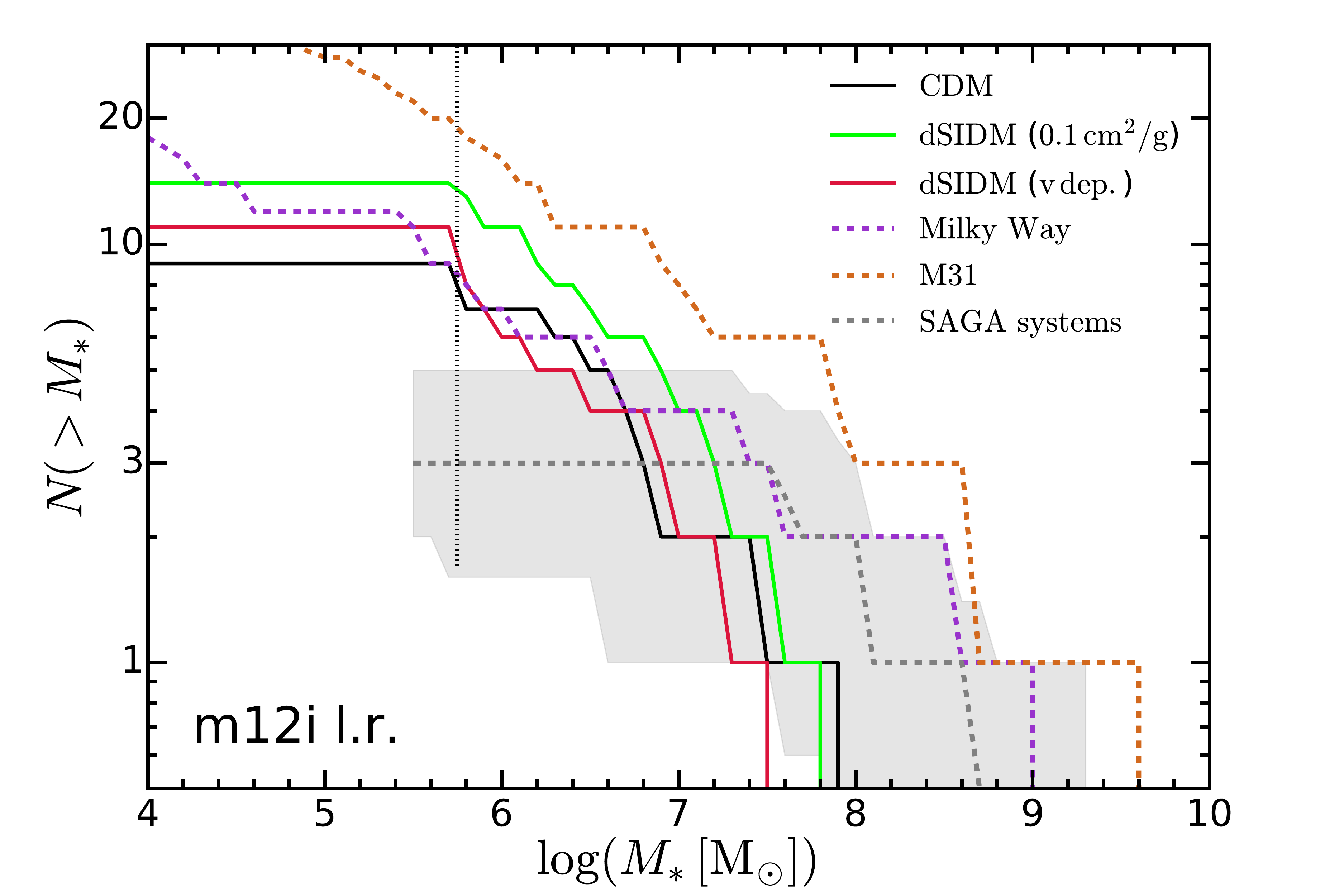}
    \includegraphics[width=0.49\textwidth]{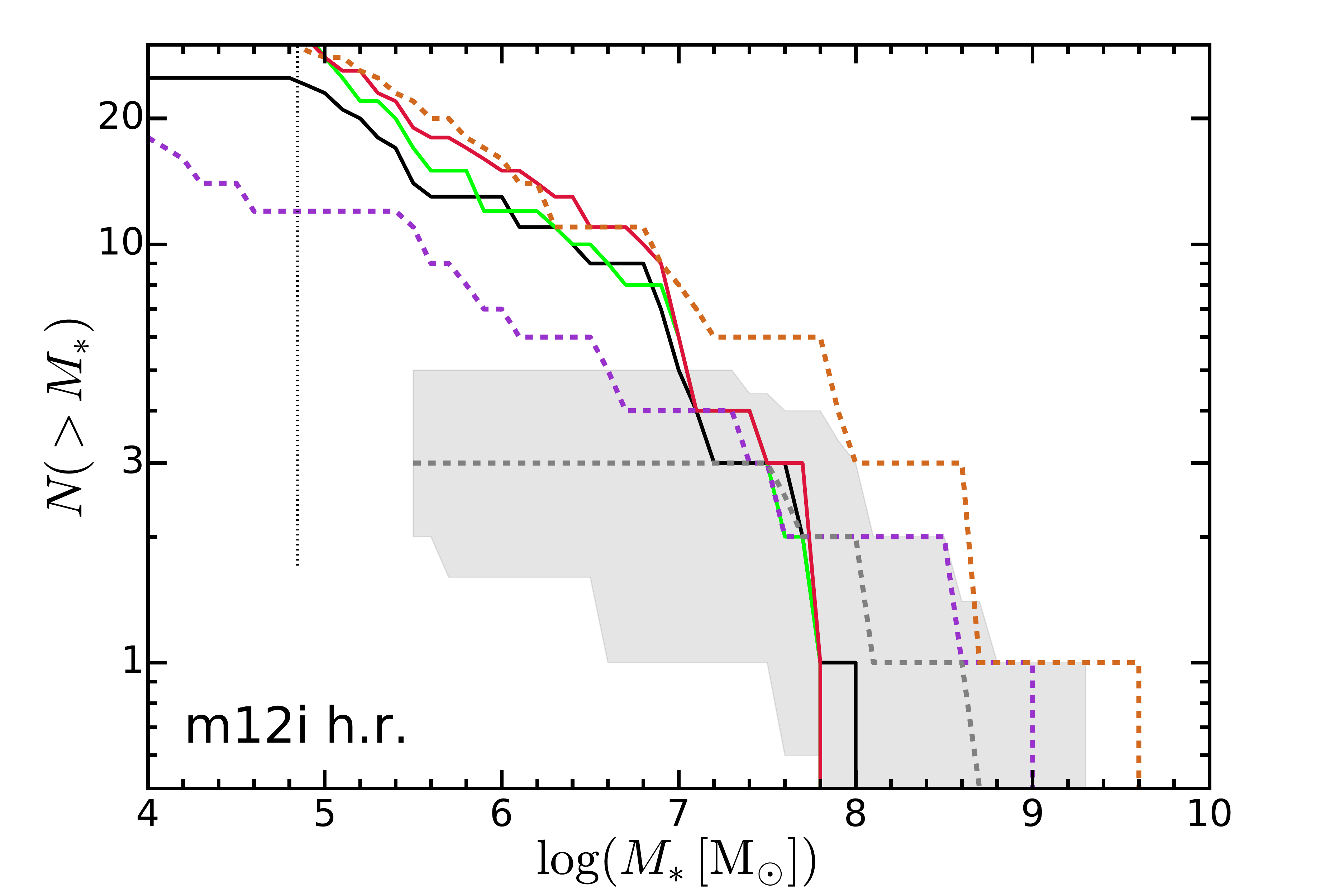}
    \includegraphics[width=0.49\textwidth]{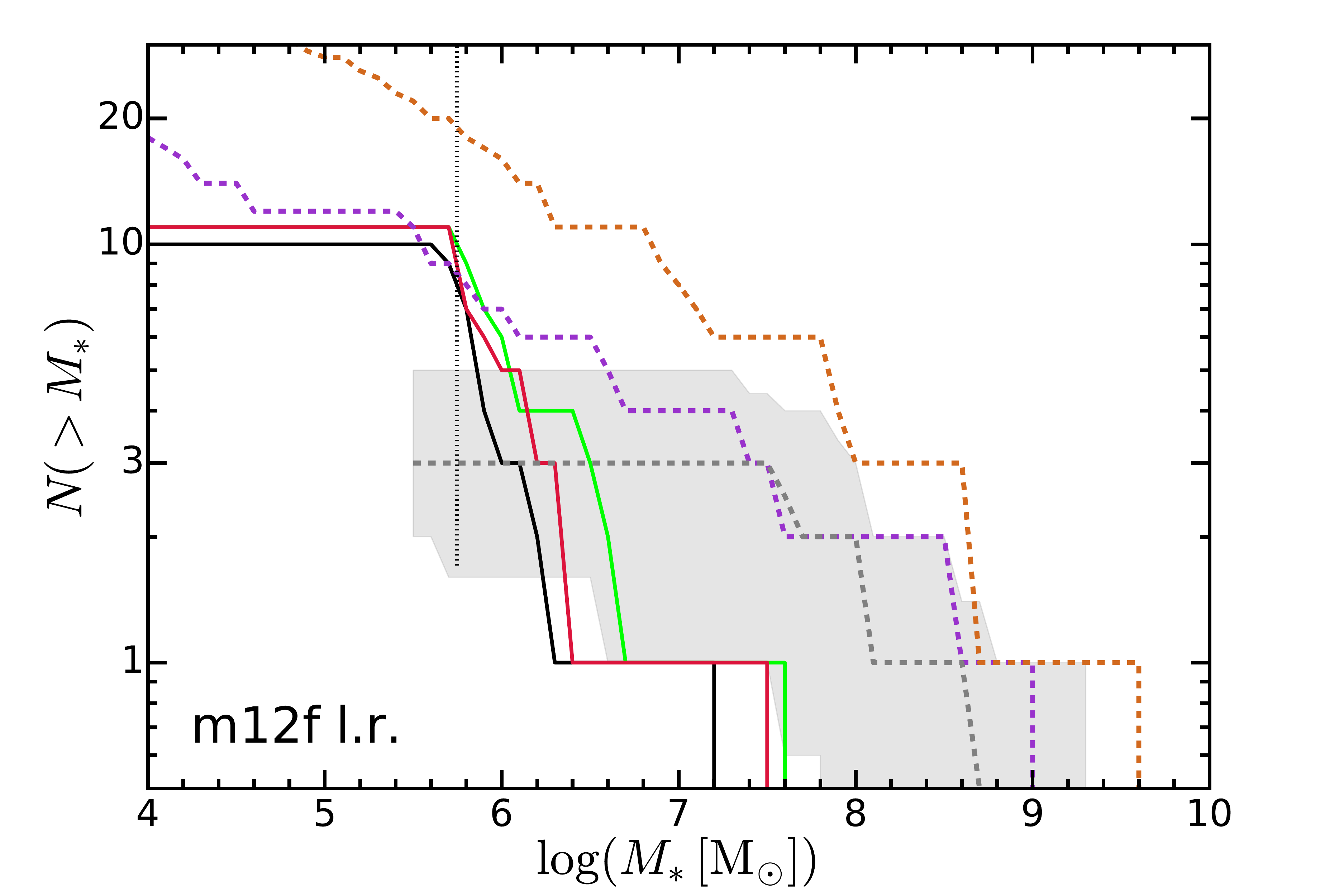}
    \includegraphics[width=0.49\textwidth]{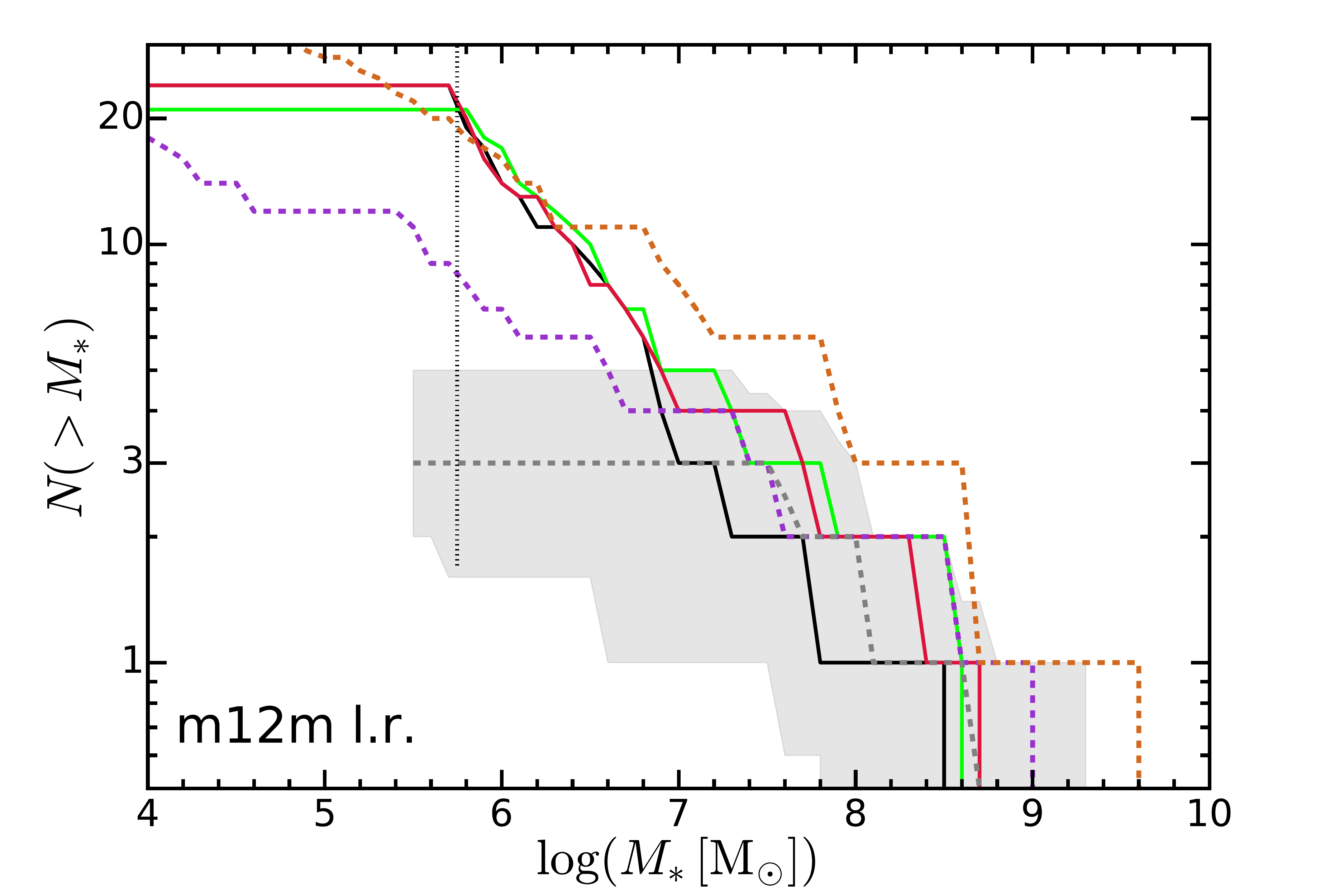}
    \caption{\textbf{Satellite stellar mass function.} The satellite stellar mass functions of different dark matter models are shown in solid lines with different colors (as labelled). The purple and orange dashed lines show the satellite stellar mass functions of the Milky Way and M31, respectively. The gray dashed lines with shaded regions show mass function of Milky Way-like systems in the SAGA survey with $1\sigma$ scatter. Each panel corresponds to one simulated Milky Way-mass galaxy in the suite. The vertical dotted line indicates the resolution limits of satellite stellar mass (set as $10$ times the baryonic mass resolution of the simulation). Strong diversity shows up in the stellar mass function of both observed satellites and the satellites of simulated galaxies. The counts of satellites get enhanced slightly in the dSIDM models, but the differences are still too small compared to the observed scatter to effectively rule out any of the model studied.}
    \label{fig:sate_massfunc}
\end{figure*}

\section{Satellite counts}
\label{sec:satellite}

In addition to the internal structure of satellites, the number counts of satellites could also serve as a channel to constrain alternative dark matter models. For example, the most well-known small-scale issue is the ``missing satellite'' (MS) problem \citep{Klypin1999,Moore1999}, which states that the dark matter subhaloes around Milky Way-mass hosts in DMO simulations outnumber the observed satellites in the actual Milky Way. The problem has been alleviated by the growing number of observed satellites in the Local Group and more realistic modelling of the baryonic physics in CDM simulations~\citep[e.g.,][]{Wetzel2016,SGK2019,Samuel2020}. 

In Figure~\ref{fig:sate_massfunc}, we show the satellite stellar mass functions from simulated Milky Way-mass galaxies and compare them to the observed mass functions of the Milky Way, M31 and $36$ Milky Way-like systems from the Exploring Satellites Around Galactic Analogs \citep[SAGA,][]{SAGA12017,SAGA22021} Survey Stage \Rmnum{2} \footnote{We acknowledge potential inconsistency in the selection criteria used between satellites in simulations and the SAGA satellites, which are selected within a line-of-sight aperture and within a line-of-sight velocity cut.}. Each panel corresponds to one of the simulations of Milky Way-mass galaxy. Following the convention in the previous section, we only select satellites with stellar mass larger than $10$ times the baryonic mass resolution of the simulations. This limit is indicated by the vertical dotted lines. For the observations, the Milky Way and M31 satellites extend to stellar mass below $10^{5}\msun$. All $36$ complete systems in SAGA reach $100\%$ spectroscopic coverage within the primary targeting region for galaxies brighter than $M_{\rm r}=-15.5$. For galaxies fainter than $M_{\rm r}=-15.5$, the survey maintain a $\sim 90\%$ spectroscopic coverage down to $M_{\rm r}=-12.3$, with completeness slightly decreasing towards fainter magnitudes. Using the color-dependent stellar mass estimates in \citet{SAGA22021} \citep[modified based on][] {Bell2003}, the limit $M_{\rm r}=-12.3$ can be translated to the stellar mass of $M_{\ast} \sim 10^{6.4-7}$ assuming the typical color $0.2 \lesssim (g-r)_{\rm 0} \lesssim 0.7$ of the confirmed satellites. This forms an estimate of the completeness limit of the SAGA surveys. The satellite mass function of simulated galaxies show significant diversity, with m12m and m12i (h.r.) hosting $\sim 10$ satellites with $M_{\ast}\gtrsim 10^{6.5}\msun$ while m12f hosts only one such satellite. This level of diversity is consistent with the scatter in mass functions revealed by the SAGA surveys. Except for m12f, which shows apparent deficiency of massive satellites, the satellite mass functions of simulated galaxies are generally consistent with observations. There are slight differences between different dark matter models. The dSIDM models with either constant or velocity-dependent cross-section do produce slightly more satellites at a given mass than CDM (i.e.\ slightly more-massive satellites by stellar mass, for a given halo mass, on average). However, the difference is subdominant compared to the scatter found in observations and none of the model tested is in tension with observations here.

\begin{figure*}
    \centering
    \includegraphics[width=0.49\textwidth]{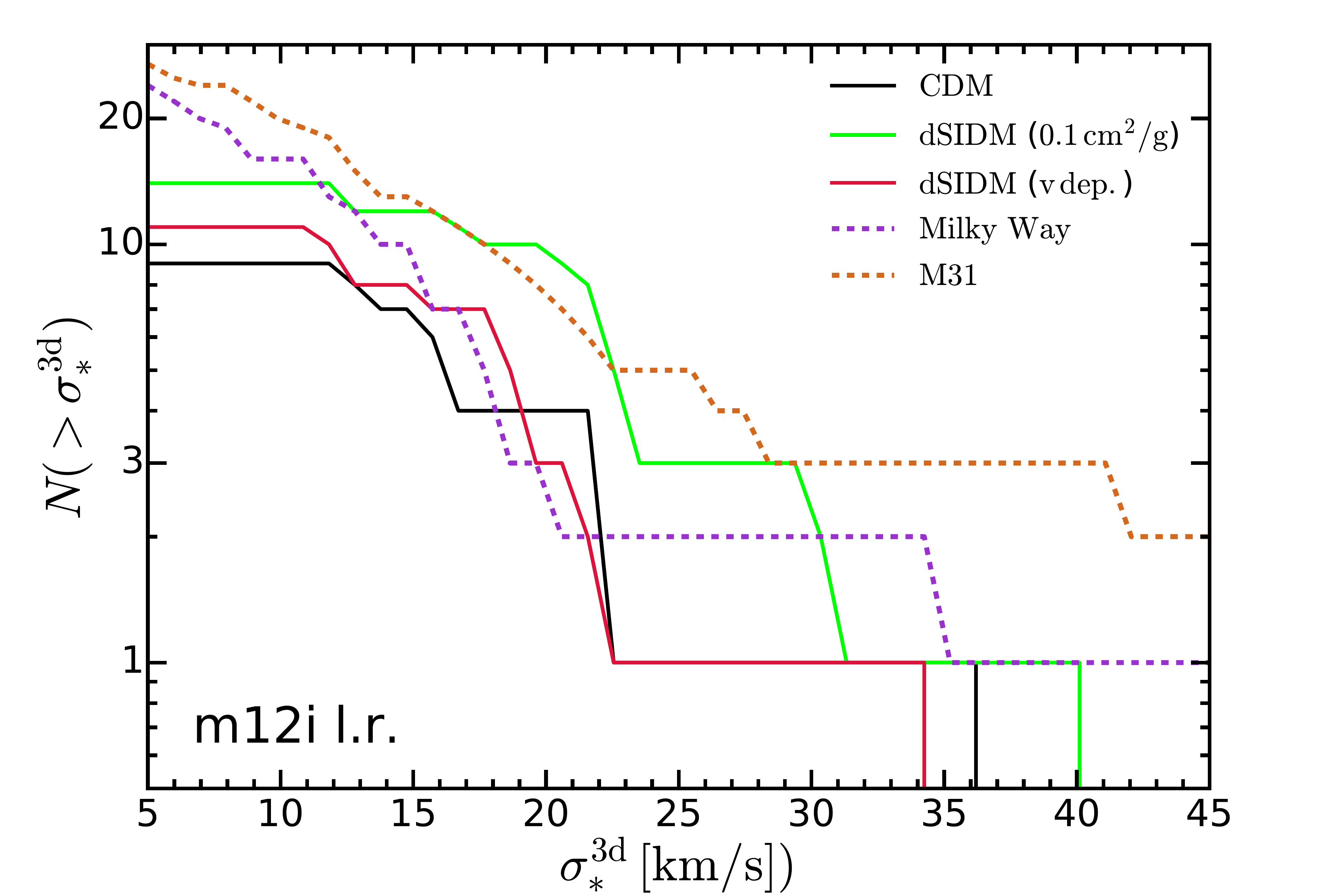}
    \includegraphics[width=0.49\textwidth]{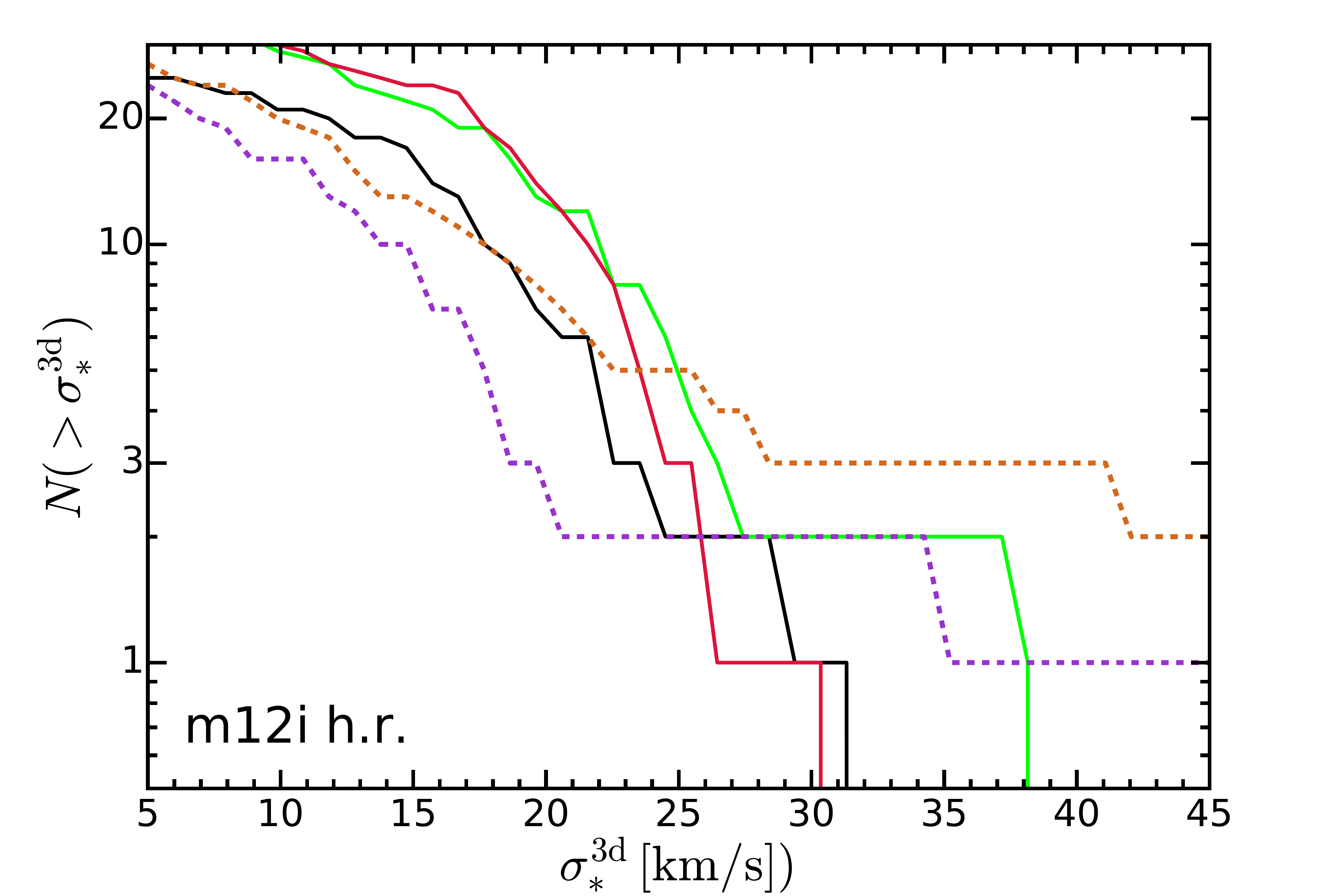}
    \includegraphics[width=0.49\textwidth]{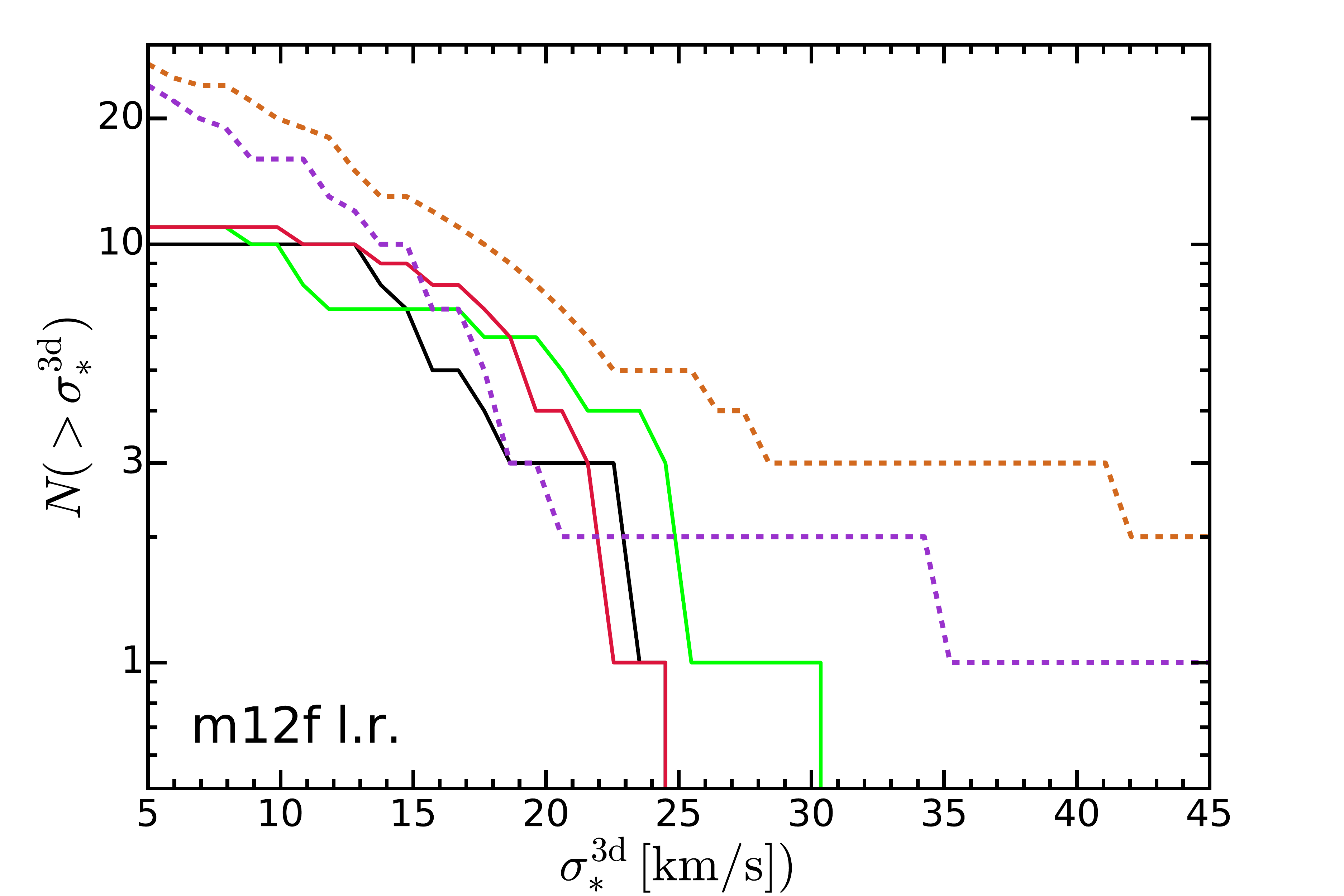}
    \includegraphics[width=0.49\textwidth]{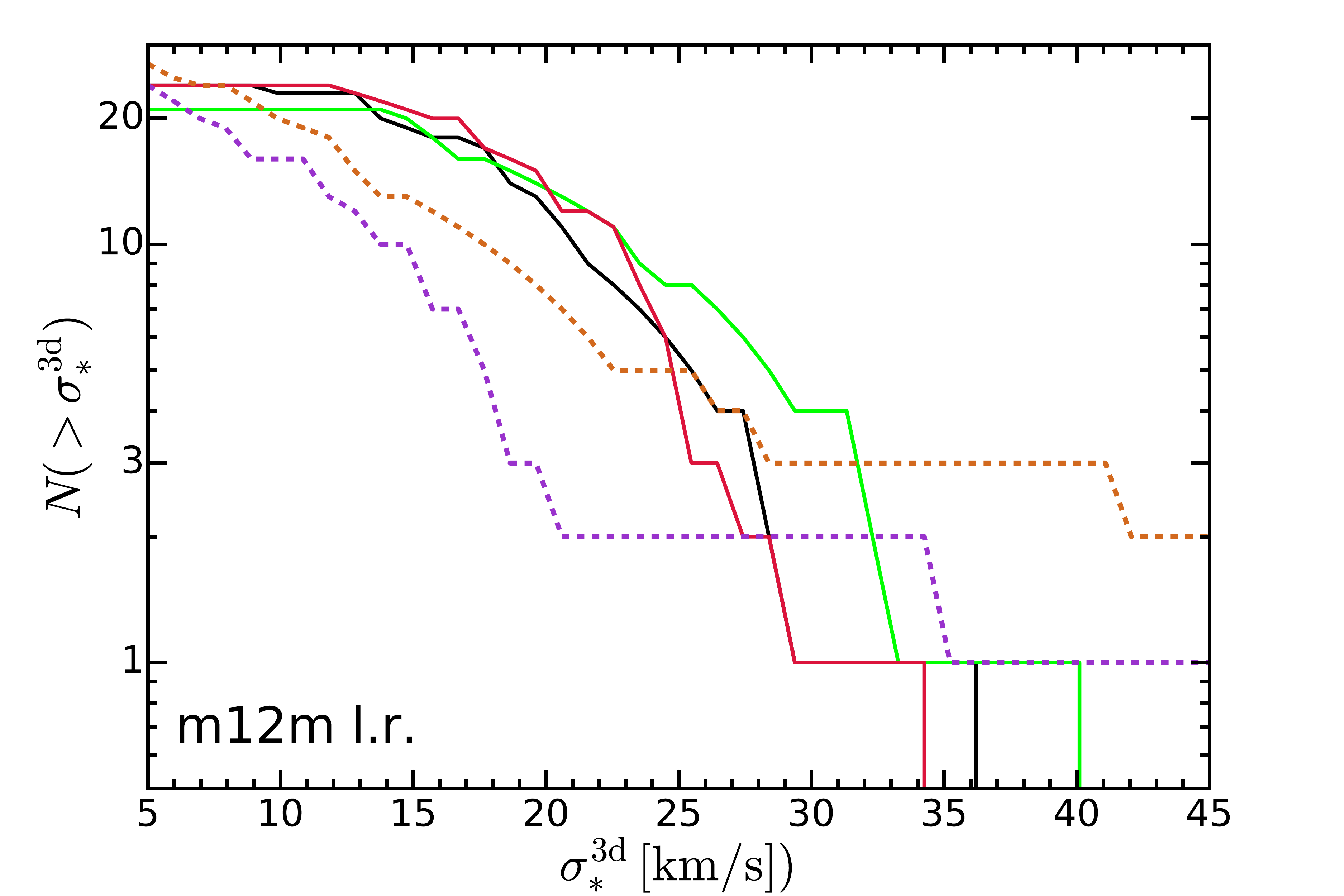}
    \caption{\textbf{Cumulative count of satellites above a given stellar 3-D velocity dispersion.} The notation is the same as Figure~\ref{fig:sate_massfunc}. Similar to the stellar mass function, we find strong diversity here in both observed and simulated systems. The satellite $\sigma^{\rm 3d}_{\ast}$ distributions of m12i (l.r.) and m12f are in good agreement with the Milky Way and M31 samples at $\sigma^{\rm 3d}_{\ast}\lesssim 20 \kms$ but do not produce enough dynamically hot satellites. On contrary, in m12i (h.r.) and m12m, the high $\sigma^{\rm 3d}_{\ast}$ end is in better agreement with the observed sample, but they tend to overpredict the number of satellites with $\sigma^{\rm 3d}_{\ast} \lesssim 25 \kms$. In terms of the dark matter physics tested, the dSIDM models (especially the dSIDM-c0.1 model) predict systematically higher velocity dispersions of satellites.}
    \label{fig:sat_vdispfunc}
\end{figure*}

In Figure~\ref{fig:sat_vdispfunc}, we show the cumulative number counts of satellites above a given stellar 3-D velocity dispersion, $\sigma^{\rm 3d}_{\ast}$. For satellites in simulations, $\sigma^{\rm 3d}_{\ast}$ is measured at $r_{1/2}$, where it is expected to reflect the total dynamical mass \citep{Walker2009}. For the observed sample, we convert the observed line-of-sight velocity dispersion to 3-D via $\sigma^{\rm 3d}_{\ast} = \sqrt{3}\,\sigma^{\rm 3d}_{\ast}$ \citep[e.g.][]{Wolf2010}. The $\sigma^{\rm 3d}_{\ast}$ distributions of m12i (l.r.) and m12f are consistent with the Milky Way and M31 satellites at $\sigma^{\rm 3d}_{\ast}\lesssim 20 \kms$ until reaching the resolution limit at low velocities. However, they do not contain as many dynamically hot satellites as the observed sample. In m12i (h.r.) and m12m, satellites exhibit systematically higher velocity dispersions (or equivalently more satellites above a given $\sigma^{3d}_{\ast}$) than m12i (l.r.) and m12f, and match better the high $\sigma^{\rm 3d}_{\ast}$ end of the observed sample. But they tend to overpredict the number of satellites with $\sigma^{\rm 3d}_{\ast} \lesssim 25 \kms$. 
The dSIDM models, especially the dSIDM-c0.1 model, produce more dynamically hot satellites in all the four Milky Way-mass galaxies simulated. This is likely caused by larger dynamical masses of the satellites at $r_{1/2}$ on average and also a few compact outliers in dSIDM-c0.1 as shown in Figure~\ref{fig:rtcurve_sate}. Although the comparisons here do not necessarily imply a particular model is favored or in tension with observations (given limited statistics of the host systems studied), it points to an interesting channel to study dissipative dark matter models.

In additional to the number count, the spatial distribution of satellites is also crucial in understanding the evolution of substructures in the Local Group environment. In particular, astrometric measurements have revealed that most of the Milky Way satellites orbit coherently within a spatially thin plane \citep[e.g.,][]{Lynden-Bell1976, Kroupa2005, Pawlowski2012} affirmed by the recent {\it Gaia} measurements \citep{Fritz2018,Pawlowski2020}. The mass and spatial distribution of satellites has been studied using FIRE-2 simulations~\citep{Samuel2020,Samuel2021} in $\Lambda$CDM. The dSIDM counterpart would be particularly interesting to explore since dissipation promotes coherent dark rotation and triggers halo deformation as found in \citealt{Shen2021}. This aspect will be investigated in follow-up papers of this series.

\section{Summary and conclusions}
\label{sec:conclusion}

This paper is the second in a series studying galaxy formation in dissipative self-interacting dark matter. In \citetalias{Shen2021}, a suite of cosmological hydrodynamical zoom-in simulations of galaxies with dSIDM was introduced. As the starting point to study structure formation in dissipative dark matter, a simplified empirical model featuring a constant fractional energy dissipation was chosen, motivated by interactions of dark matter composites (for example, confined particles in a non-Abelian hidden sector or large stable bound states (dark ``nuggets'') of asymmetric dark matter). Several interesting phenomena and physics on the dark matter side, related to dSIDM, were identified in \citetalias{Shen2021}.

In this paper, we attempt to compare predictions to basic galaxy observables affected by the underlying structural changes of dark matter haloes induced by dissipative interactions. The stellar morphology, the size-mass relation and the circular velocity profiles of both field and satellite dwarf galaxies are studied, and first constraints on the dSIDM model are obtained through comparisons with observations of local dwarf galaxies. 

We first study the observed morphology of the stellar component and quantitatively the size-mass relation of isolated dwarf galaxies. With moderate but not negligible interaction cross-sections ($(\sigma/m) \sim 1\cpm$), dSIDM makes the stellar content more concentrated and promotes the formation of thin stellar disks as well as neutral gas disks in massive bright dwarfs. The simulated galaxies in these models are still consistent with observations in the plane of the galaxy size-mass relation. However, perhaps surprisingly, when the cross-section becomes large enough ($ \sigma/m \sim 10\cpm$), the stellar content of simulated dwarfs becomes fluffier even than the CDM case, owing to rotation and other emergent properties of the dark matter cusp. The dwarfs in this model lie systematically at the most diffuse observed end of the size-mass relation and thus this model faces strong constraints. 

In terms of the circular velocity profiles of simulated dwarfs, we separately consider the isolated classical and bright dwarfs in the suite as well as the satellites in the simulations of Milky Way-mass galaxies. The isolated classical dwarfs are compared to the field dwarf galaxies in the Local Group and we find all of the dSIDM models studied survive this comparison. The isolated bright dwarfs are compared to the LSBs with H\Rmnum{1}-based circular velocity measurements. We find that the dSIDM models with $(\sigma/m)\gtrsim 0.1\cpm$ are in tension with observations and the velocity-dependent model is favored. The satellites in simulated Milky Way-mass galaxies are compared to the Local Group satellites. Though we find little differences in the median and scatter of the circular velocity profiles between dark matter models, dSIDM models with $(\sigma/m) = 0.1\cpm$ produce outliers that agree better with the compact elliptical satellites in observations, whose analogs are missing in CDM. Although the circular velocity profiles of satellites in simulations are consistent with the observationally inferred velocity dispersions of these systems, the size of the simulated satellites are systematically larger. However, this is potentially subjected to selection bias in observations and also could be a resolution effect. Further high-resolution simulations are required to resolved the central kinematic structure of satellites to give more robust predictions. Meanwhile, the stellar mass function and velocity dispersion function of satellites are studied. In dSIDM models, the number count of satellite galaxies is slightly enhanced and the satellites are dynamically hotter, but the difference is too small to infer valid constraints on the models. 

In conclusion, it is at the mass scale of isolated bright dwarfs that the dSIDM models with constant cross-sections face the most stringent constraint, and models with $(\sigma/m) \gtrsim 0.1\cpm$ are in tension with H\Rmnum{1}-based circular velocity measurements. The constraint is much weaker in lower-mass isolated dwarfs or in satellites of Milky Way-mass hosts. Since as shown in \citetalias{Shen2021} the dSIDM-related phenomena strictly depend on the dissipation time scale, which is inversely proportional to the product of $f_{\rm diss}$ and $(\sigma/m)$, the constraints derive here can be translated to other $f_{\rm diss}$ values giving the combined constraints: $f_{\rm diss}\, (\sigma/m) \lesssim 0.05\cpm$. In future work, it would be helpful to improve the robustness of the constraints here with better statistics of simulations (simulating a greater variety of galaxies). Of course improved resolution would help to resolve the central structure of satellite galaxies, and in particular to investigate the implication of dSIDM in explaining the diversity of dwarf compactness in the Local Group.

%%%%%%%%%%%%%%%%%%%%%%%%%%%%%%%%%%%%%%%%%%%%%%%%%%

\section*{Acknowledgements}

Support for XS \&\ PFH was provided by the National Science Foundation (NSF) Research Grants 1911233, 20009234, 2108318, the NSF Faculty Early Career Development Program (CAREER) grant 1455342, the National Aeronautics and Space Administration (NASA) grants 80NSSC18K0562, HST-AR-15800. Numerical calculations were run on the supercomputer Frontera at the Texas Advanced Computing Center (TACC) under the allocations AST21010 and AST20016 supported by the NSF and TACC, and NASA HEC SMD-16-7592.
FJ is supported by the Troesh scholarship.
MBK acknowledges support from NSF CAREER award AST-1752913, NSF grants AST-1910346 and AST-2108962, NASA grant NNX17AG29G, and HST-AR-15006, HST-AR-15809, HST-GO-15658, HST-GO-15901, HST-GO-15902, HST-AR-16159, and HST-GO-16226 from the Space Telescope Science Institute (STScI), which is operated by AURA, Inc., under NASA contract NAS5-26555.
AW received support from: NSF grants CAREER 2045928 and 2107772; NASA  Astrophysics Theory Program (ATP) grant 80NSSC20K0513; HST grants AR-15809, GO-15902, GO-16273 from STScI.
This research made use of data from the SAGA Survey (sagasurvey.org). The SAGA Survey was supported by NSF collaborative grants AST-1517148 and AST-1517422 and by Heising–Simons Foundation grant 2019-1402.

\section*{Data Availability}
The simulation data of this work was generated and stored on the supercomputing system \href{https://frontera-portal.tacc.utexas.edu/about/}{Frontera} at the Texas Advanced Computing Center (TACC), under the allocations AST20010/AST20016 supported by the NSF and TACC, and NASA HEC SMD-16-7592. The CDM FIRE-2 simulations are publicly available \citep{Wetzel2022} at \url{http://flathub.flatironinstitute.org/fire}. However, the data of the dSIDM simulations used in this article cannot be shared publicly immediately, since the series of paper is still in development. The data will be shared on reasonable request to the corresponding author.



%\bibliographystyle{mnras}
%\bibliography{reference} % if your bibtex file is called example.bib


%%%%%%%%%%%%%%%%% APPENDICES %%%%%%%%%%%%%%%%%%%%%

\appendix
\section{Resolution dependence of satellite properties}
\label{app:resolution}

The analysis above utilizes both low and high-resolution Milky Way-mass galaxies in the simulation suite. However, the satellite structure could be resolution-dependent. This can arise from two primary causes: (\rmnum{1}), the N-body relaxation of collisionless particles; and (\rmnum{2}), the artificial burstiness of the star formation history due to limited mass resolution (discreteness effects). Both can puff up the dark matter and the stellar content of low-mass galaxies artificially. For example, in \citet{Fitts2019}, the test on the isolated classical dwarf m10b has shown that the $r_{\rm 1/2}$ shrinks by about a factor of two (despite minimal changes of the overall halo properties) when increasing the mass resolution from $m_{\rm b}=4000\msun$ to $m_{\rm b}=62.5\msun$. Similar resolution effects manifested in the comparison of the observed ultra-faint dwarfs with high-resolution dwarf simulations in \citet{Wheeler2019}.

\begin{figure}
    \centering
    \includegraphics[width=0.49\textwidth]{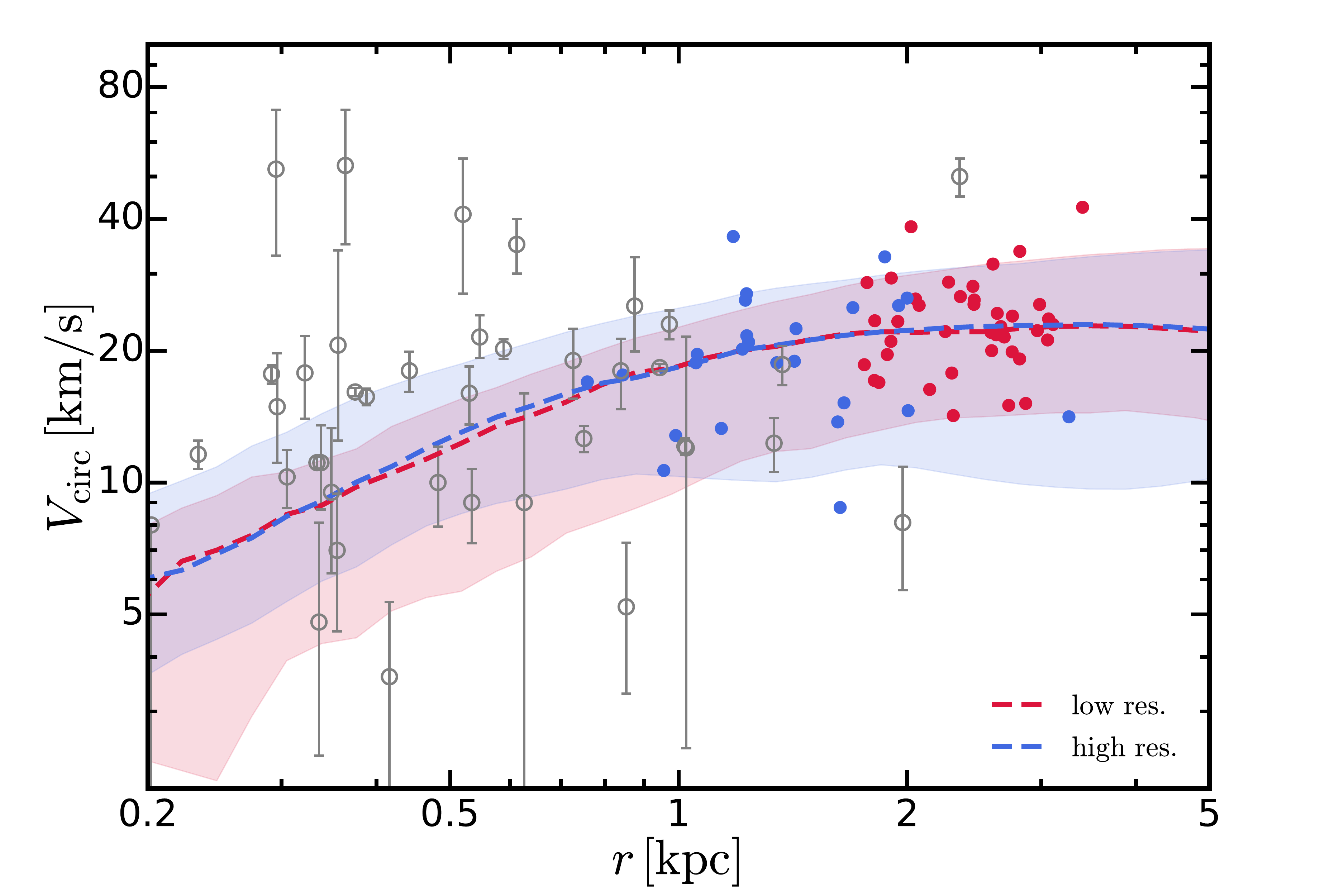}
    \caption{We compare satellite circular velocity profiles from the high and low-resolution simulations of m12i. The notation follows the top panel. Although the median circular velocity profile and the scatter do not differ appreciably between high and low-resolution simulations, the $r_{\rm 1/2}$'s of satellites in simulations are systematically smaller in the high-resolution simulation. Compared to the observed dwarfs, even the high-resolution simulation produces fluffier stellar content for these satellites. }
    \label{fig:sat_res}
\end{figure}

In Figure~\ref{fig:sat_res}, we compare the satellite circular velocity profiles from the high and low-resolution simulations of m12i (listed in Table~\ref{tab:sim}). Aside from the median and scatter of circular velocity profiles, we also show the $(V_{\rm 1/2},\, r_{\rm 1/2})$ of these satellites. The median circular velocity profile is converged and the $1.5\sigma$ ($7\%$ to $93\%$ inclusion) contour moves up slightly. This indicates that the underlying dark matter structure of these satellites is converged at the resolution level. However, the $r_{\rm 1/2}$'s are systematically smaller in the high-resolution run and the factor of by which they change is consistent with the enhancement in spatial resolution (two times higher spatial resolution and eight times better mass resolution). Even in the high-resolution run (the mass resolution of which is still at least an order of magnitude poorer than that of isolated dwarf galaxy simulations), the stellar content of satellites can only be resolved to about $1\kpc$ scale, and so the simulated small satellites are more extended than the observed satellites. 

\begin{figure}
    \centering
    \includegraphics[width=0.49\textwidth]{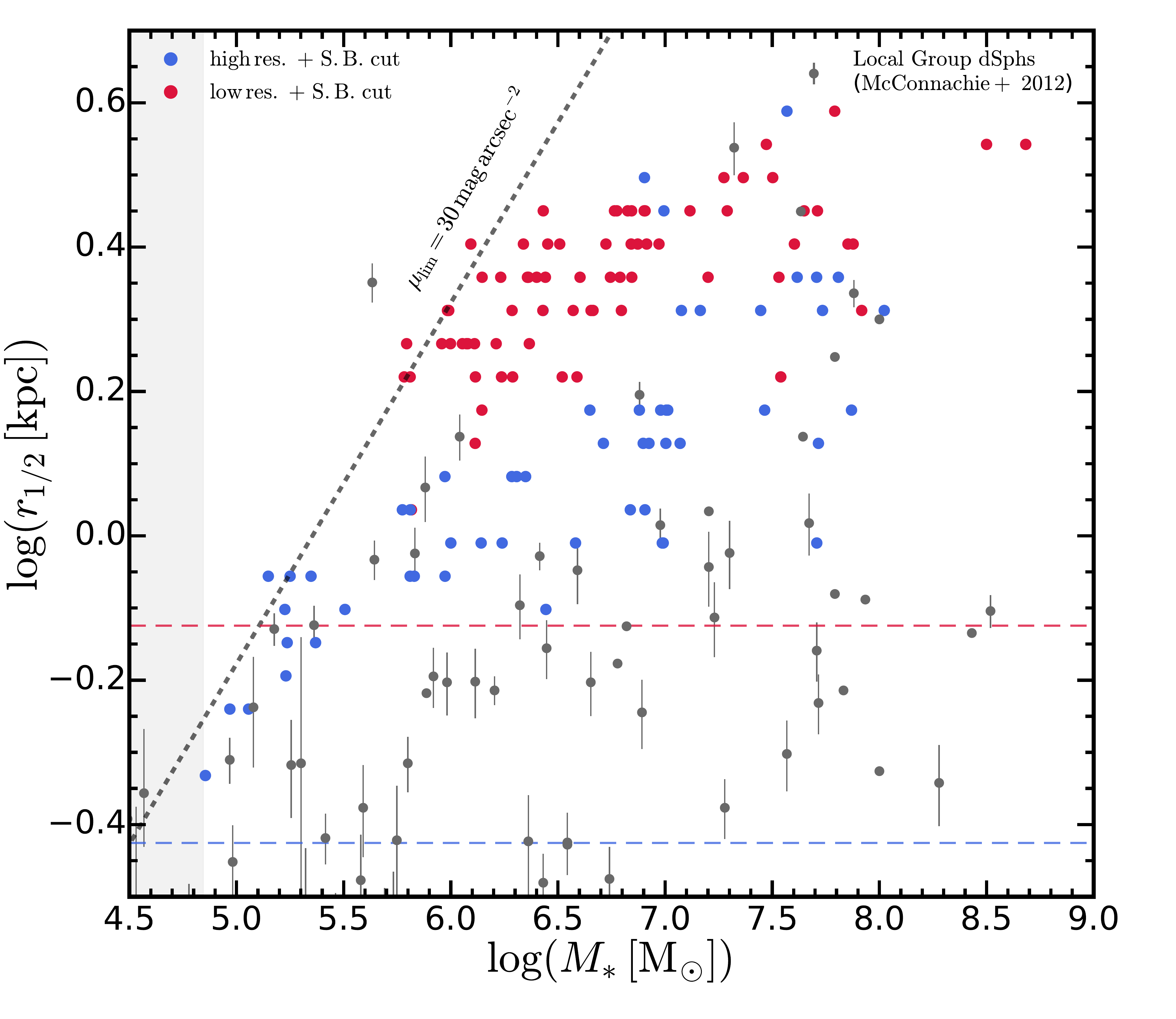}
    \caption{We compare the high- and low-resolution simulations on the plane of the size-mass relation. They are all corrected for the surface brightness limit at $\mu_{\rm V} = 30\mmag\,{\rm arcsec}^{-2}$. The horizontal lines indicate the radius enclosing $200$ dark matter particles assuming the typical satellite central density $\rho_{\rm dm} = 10^{7.5}\msun\,\kpc^{-3}$. Satellites in the low-resolution simulations are systematically more diffuse than their high-resolution counterparts. The resolution dependence could explain the discrepancy of the simulations with observations in this plane.}
    \label{fig:sat_res_size}
\end{figure}

In Figure~\ref{fig:sat_res_size}, we show the size-mass relations of satellites from the high- and low-resolution simulations. The satellite stellar mass and size have been corrected for the surface brightness cut-off at $\mu_{\rm V} = 30 \mmag\, {\rm arcsec}^{-2}$. The satellite sizes in the low-resolution runs are systematically higher than the high-resolution ones. The horizontal lines indicate the radius enclosing $200$ dark matter particles assuming the typical satellite central density $\rho_{\rm dm} = 10^{7.5}\msun\,\kpc^{-3}$. The number $200$ is suggested in \citet{Hopkins2018} as the convergence criterion in dark matter properties for FIRE-2 simulations. This limit roughly gives the minimum $r_{1/2}$ that the simulation can resolve. Certainly, we cannot conclude that the satellite sizes are fully resolved even in the high-resolution runs, and it is likely that increasing the resolution will give better agreement with the observed satellites. This is supported by that the simulated isolated dwarfs in the mass range $10^{5}\operatorname{-}10^{6}\msun$ (with baryonic mass resolution $\sim 250 \operatorname{-} 500\msun$) agree well with the observations on the size-mass plane as shown in Figure~\ref{fig:mass-size} and Figure~\ref{fig:sat_rhalf}. The impact of resolution on satellite properties of Milky Way-mass hosts will be explored more in the upcoming Triple {\it Latte} simulations (with baryonic mass resolution $\sim 880 \msun$) (Wetzel et al. in prep).

%%%%%%%%%%%%%%%%%%%%%%%%%%%%%%%%%%%%%%%%%%%%%%%%%%

% Don't change these lines
\bsp	% typesetting comment
\label{lastpage}
\end{document}